\documentclass[twocolumn]{aa}
\usepackage{graphicx}
\usepackage{array}
\usepackage{amsfonts}
\usepackage{natbib}

\def\etal{{\it et al.}}
\def\llabel#1{\label{#1}}

\def\eps{\varepsilon}

\def\m@th{\mathsurround=0pt}
\def\EQM#1{\vcenter{\normalbaselines\m@th
    \ialign{${\displaystyle ##}$\hfil&&\ ${\displaystyle ##}$\hfil\crcr
    \mathstrut\crcr\noalign{\kern-\baselineskip}
    \noalign{\smallskip}
    #1\crcr\mathstrut\crcr\noalign{\kern-\baselineskip}}}}

\def\eps{\varepsilon}
\def\etal{{\it et al.}}

\newcommand{\Frac}[2]{{{\displaystyle\strut#1}\over{\displaystyle\strut#2}}}

\newcommand\be{\begin{equation}}
\newcommand\ee{\end{equation}}
\def\espace{\rule[-3pt]{0pt}{15 pt} }

\def\dt{\Delta t}

\def\pam{\phantom{$-$0}}
\def\pa{\phantom{0}}
\def\phm{\phantom{$-$}}

\def\figuresN{figs}
\def\f_eps{}
\def\f_ps{}
\def\jxlnull#1{}

\newcommand\figp{
\begin{figure}
 \includegraphics[scale=0.5]{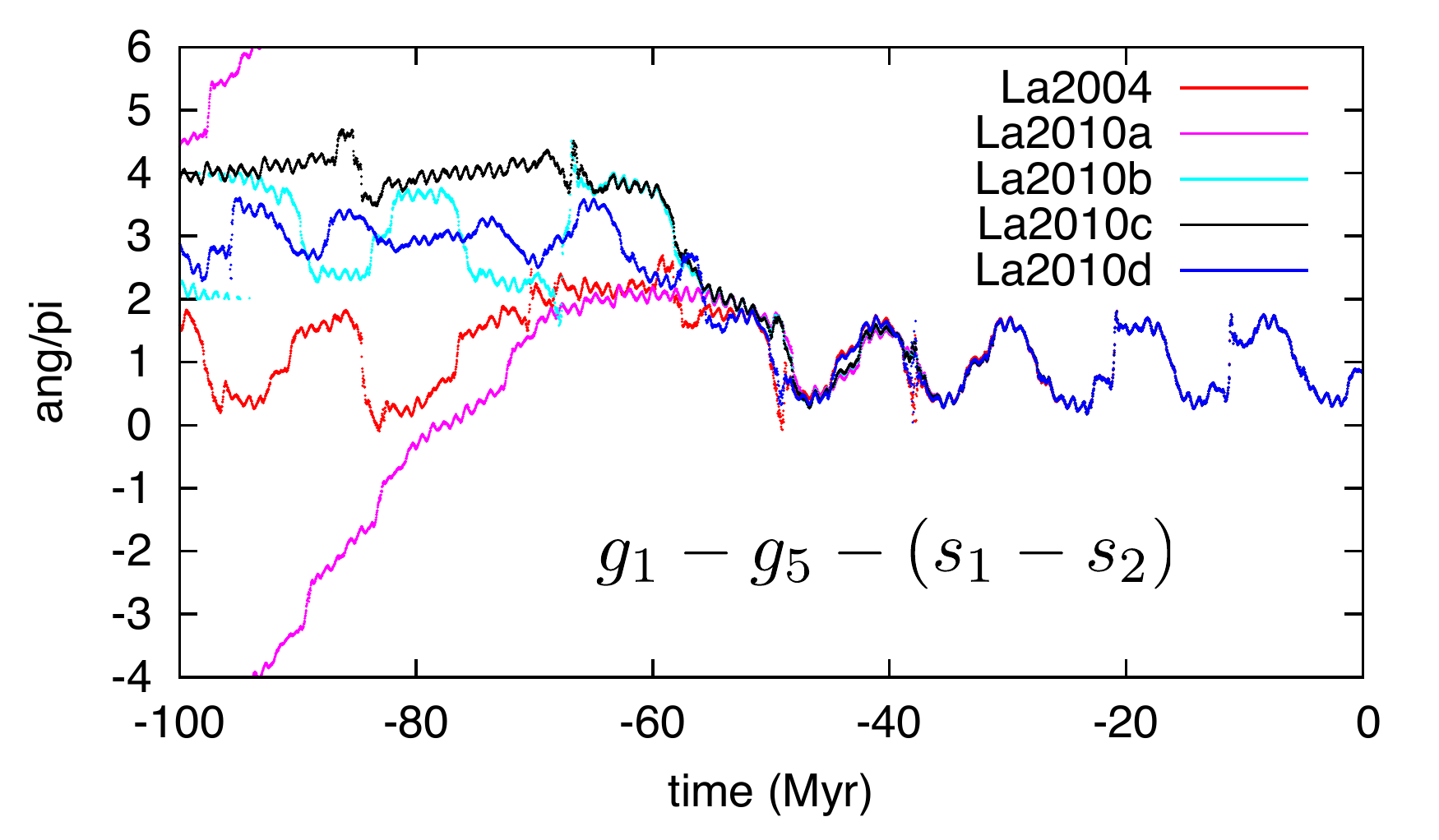} 
 \includegraphics[scale=0.5]{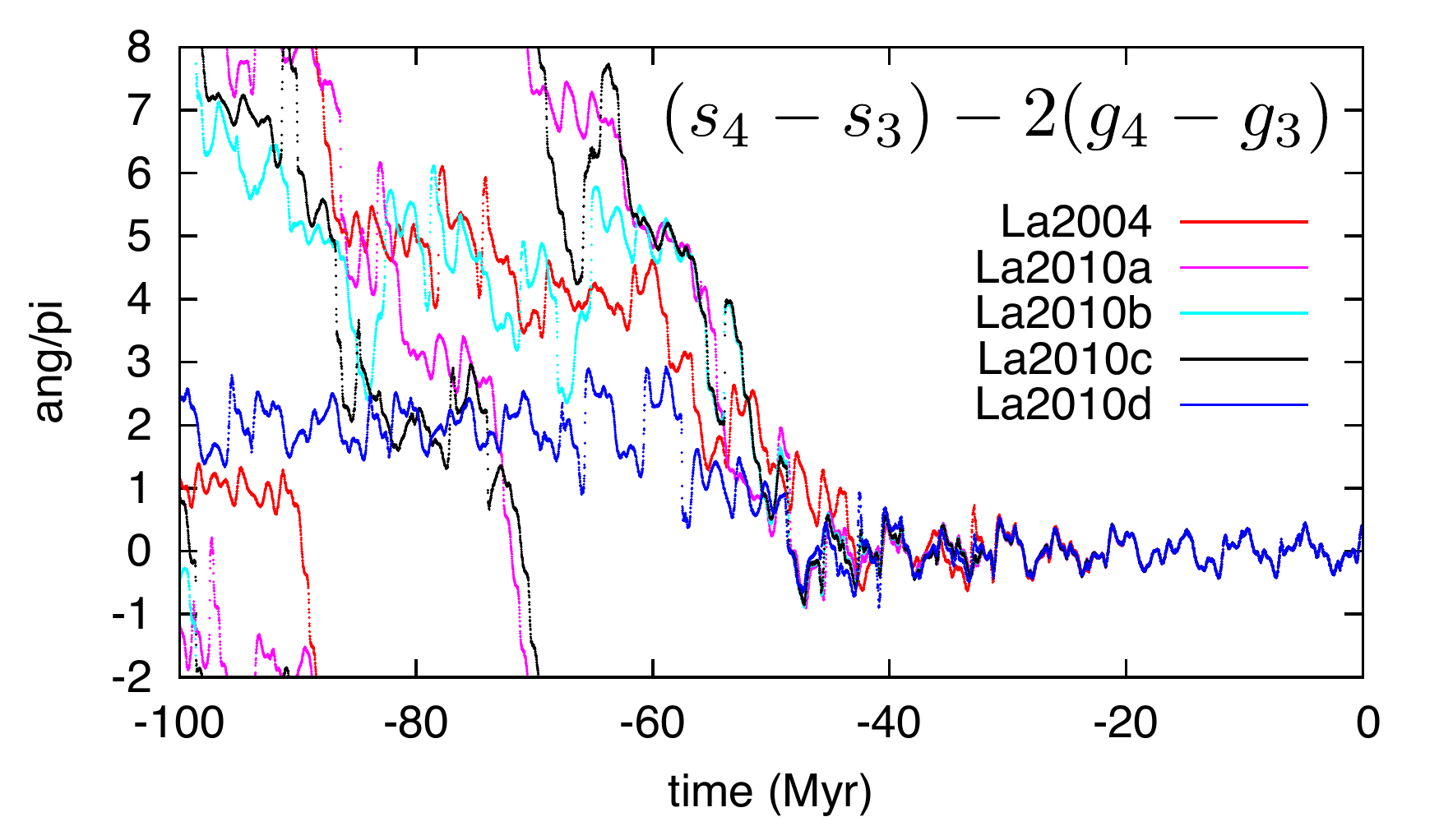} 
 \includegraphics[scale=0.5]{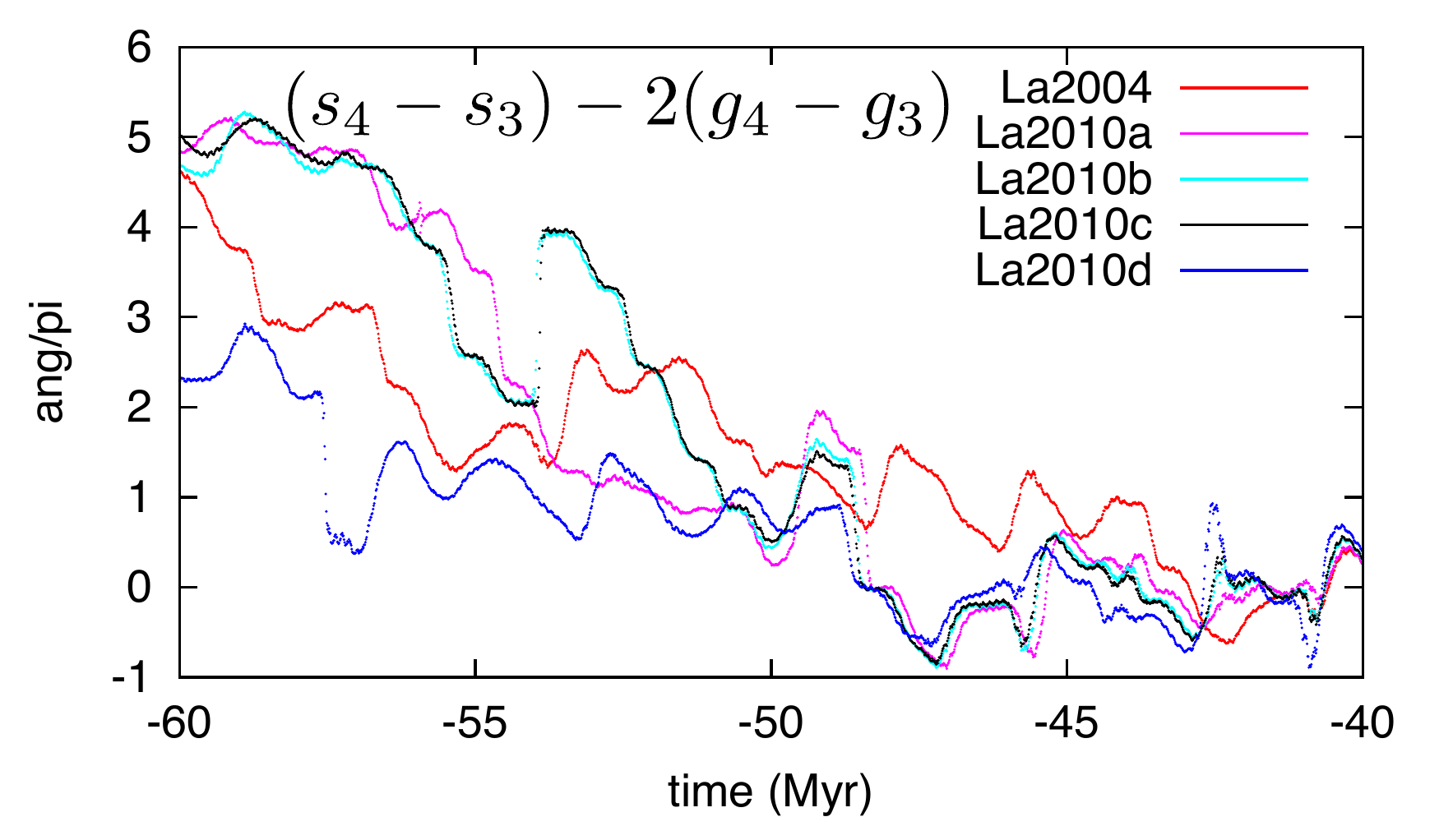} 
  \caption{Resonant arguments (in radians versus time in Myr) 
  $\sigma=(g_1-g_5)-(s_1-s_2)$ (top) and
  $\theta= (s_4-s_3)-2(g_4-g_3)$ (middle and bottom)   
 for the  different solutions La2004, La2010a,b,c,d.
 In the bottom plot, $\theta$ is diplayed at a greater scale.
 One can see that the La2010 solutions are very close up to 50 Myr.
} 
 \llabel{Figp}
\end{figure}
}

\newcommand\figr{
\begin{figure}
 \includegraphics[scale=0.5]{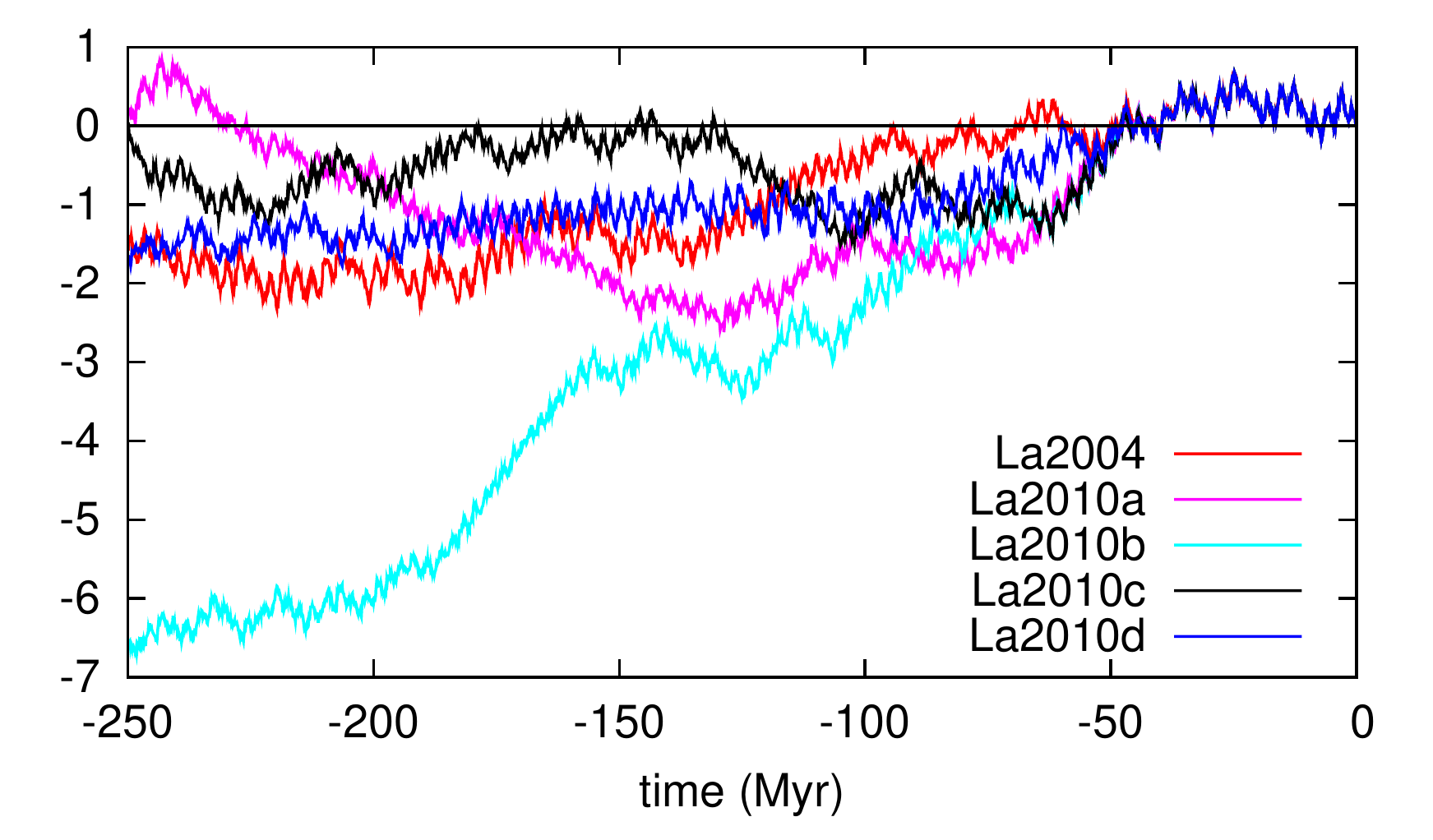} 
  \caption{405 kyr term in eccentricity.
Maximum  difference (in radians) of the argument $\theta_{g_2-g_5}(t) - \theta_{g_2-g_5}(0) $ of $g_2-g_5$ 
in all solutions La2004, La2010a,b,c,d 
with respect to the linear approximation 
$
\theta_0(t)= 3.200''\, t
$
where $t$ is in yr.
  } 
 \llabel{Figr}
\end{figure}
}

\jxlnull{

}

\newcommand\tabc{
\begin{table} 
\caption{Main secular frequencies $g_i$ and $s_i$ of La2004 and La2010a determined over 20 Ma for the four inner 
planets, and over 50 Ma for the 5 outer planets (in arcsec yr$^{-1}$).  $\Delta_{100}$   
are the observed variations of the frequencies over  100 Myr \citep{LaskRobu2004a}. In the last column, the period 
of the secular term are  given.}
\begin{tabular}{rlllr}
\hline
    &  La2004     & La2010a            &  $\Delta_{100}$  &period (yr) \\ 
\hline
$g_1$ & \pam     5.59        & \pam     5.59      &  0.13     &   231\,843 \\
$g_2$ & \pam     7.452       & \pam     7.453     &  0.019    &   173\,913 \\
$g_3$ & \phm     17.368      & \phm    17.368     &  0.20     &    74\,620 \\
$g_4$ & \phm     17.916      & \phm    17.916     &  0.20     &    72\,338 \\
$g_5$ & \pam     4.257452    & \pam     4.257482 &  0.000030  &   304\,407 \\
$g_6$ & \phm     28.2450     & \phm    28.2449 &  0.0010      &    45\,884 \\
$g_7$ & \pam     3.087951    & \pam     3.087946 &  0.000034  &   419\,696  \\
$g_8$ & \pam     0.673021    & \pam     0.673019 &  0.000015  &  1\,925\,646  \\
$g_9$ & \pa     $-$0.34994   & \pa     -0.35007 &  0.00063    &  3\,703\,492   \\
\hline
$s_1$ &\pa     $-$5.59       &\pa     -5.61   &   0.15           &    231\,843 \\
$s_2$ &\pa     $-$7.05       &\pa     -7.06   &   0.19           &    183\,830 \\
$s_3$ &       $-$18.850      &       -18.848   &   0.066         &     68\,753 \\
$s_4$ &       $-$17.755      &       -17.751   &   0.064         &     72\,994 \\
$s_5$ & \pam        &     &      &     \\
$s_6$ &       $-$26.347855   &       -26.347841 &   0.000076     &     49\,188 \\ 
$s_7$ & \pa    $-$2.9925259  & \pa    -2.9925258 &   0.000025     &    433\,079 \\
$s_8$ & \pa    $-$0.691736   & \pa    -0.691740 &   0.000010     &  1\,873\,547 \\
$s_9$ & \pa    $-$0.34998    & \pa    -0.35000 &   0.00051       &   3\,703\,069 \\ 
\hline
\end{tabular}
\llabel{tab.freqsec}
\end{table}
}

\newcommand\tabd{
\begin{table} 
\caption{Frequency decomposition of $z=e\exp i\varpi$ for the Earth on the 
time interval $[-15,+5]$ Myr (Laskar \etal, 2004).}
\begin{tabular}{rcrrr}
\hline
   $n$ &    &   $\mu_k$ (''/yr) &    $b_k$ & $\varphi_k$ (degree) \\ 
\hline
  1 &$g_5$ &    4.257564 &     0.018986 & $      30.739$\\  
  2 &$g_2$  &    7.456665 &     0.016354 & $    -157.801$\\  
  3 &$g_4$  &   17.910194 &     0.013055 & $     140.577$\\  
  4 &$g_3$  &   17.366595 &     0.008849 & $     -55.885$\\  
  5 &$g_1$  &    5.579378 &     0.004248 & $      77.107$\\  
\hline
\end{tabular}
\llabel{tab.zecc}
\end{table}
}

\def\fign{figs}

\newcommand\figNa{
\begin{figure}
 \includegraphics[scale=0.45]{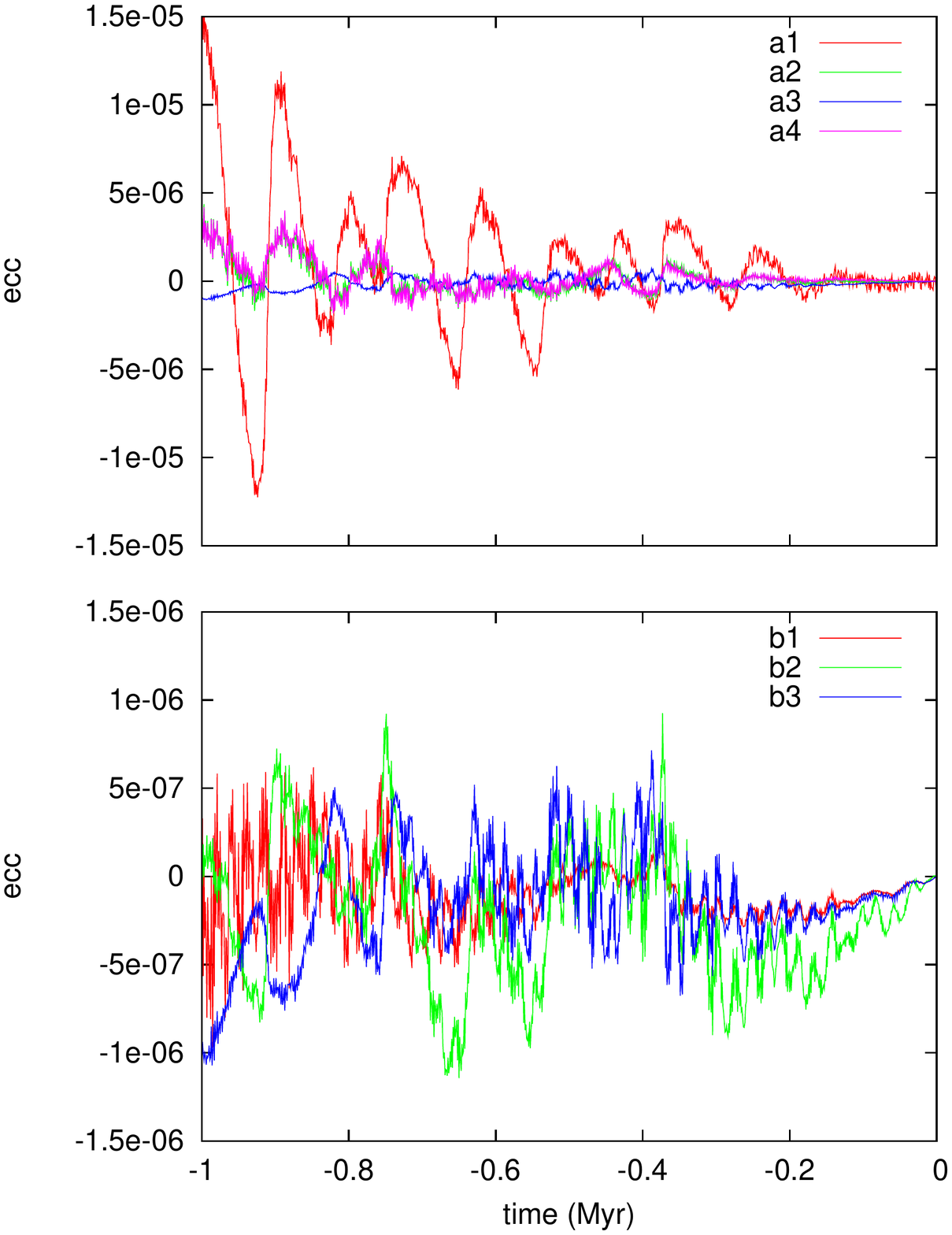} 
  \caption{
  Differences in the eccentricity of the Earth Moon barycenter over 1 Myr for various 
  solutions as follows : a1 = La2004-INPOP08; a2  = La2010d-INPOP08; a3 =b3= La2010a - INPOP08;
  a4=INPOP06-INPOP08;
  b1 = La2010b - INPOP08; b2 = La2010c -INPOP06.
  It should be noted that the vertical scale is enlarged 10 times in the bottom plot.
  All solutions are compared to INPOP08. On the top figure, La2010d (a2) and INPOP06 (a4) 
  are almost superposed. This is because La2010d was adjusted over INPOP06.
  } 
 \llabel{FigNa}
\end{figure}
}

\newcommand\figNb{
\begin{figure*}[t]
 \includegraphics[width=17cm]{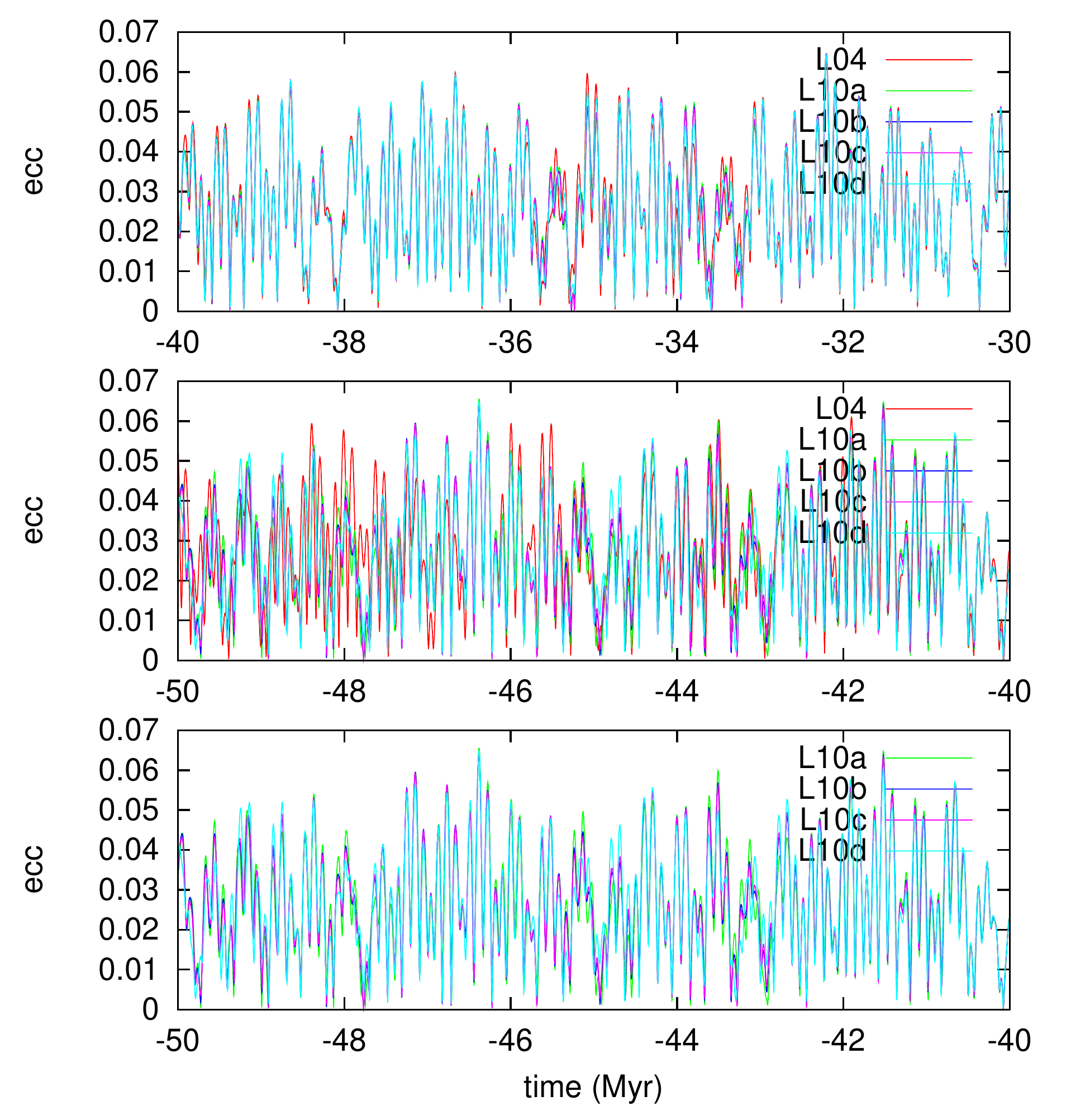} 
  \caption{
  Eccentricity of the Earth for the solutions La2004 (L04), La2010a (L10a),
  La2010b (L10b), La2010c (L10c), and La2010d (L10d).
  On the interval $[-40:-30]$  Myr, the four solutions are virtually identical, 
  but before $-45$ Myr, La2004 begins to depart significantly from the La2010 solutions.
  We have then plotted again the time interval $[-50:-40]$ Myr (bottom), removing La2004
  for a better comparison of the La2010 solutions. Over this interval, these three solutions 
  are very similar.
    } 
 \llabel{FigNb}
\end{figure*}
}

\newcommand\figNc{
\begin{figure*}[t]
 \includegraphics[width=17cm]{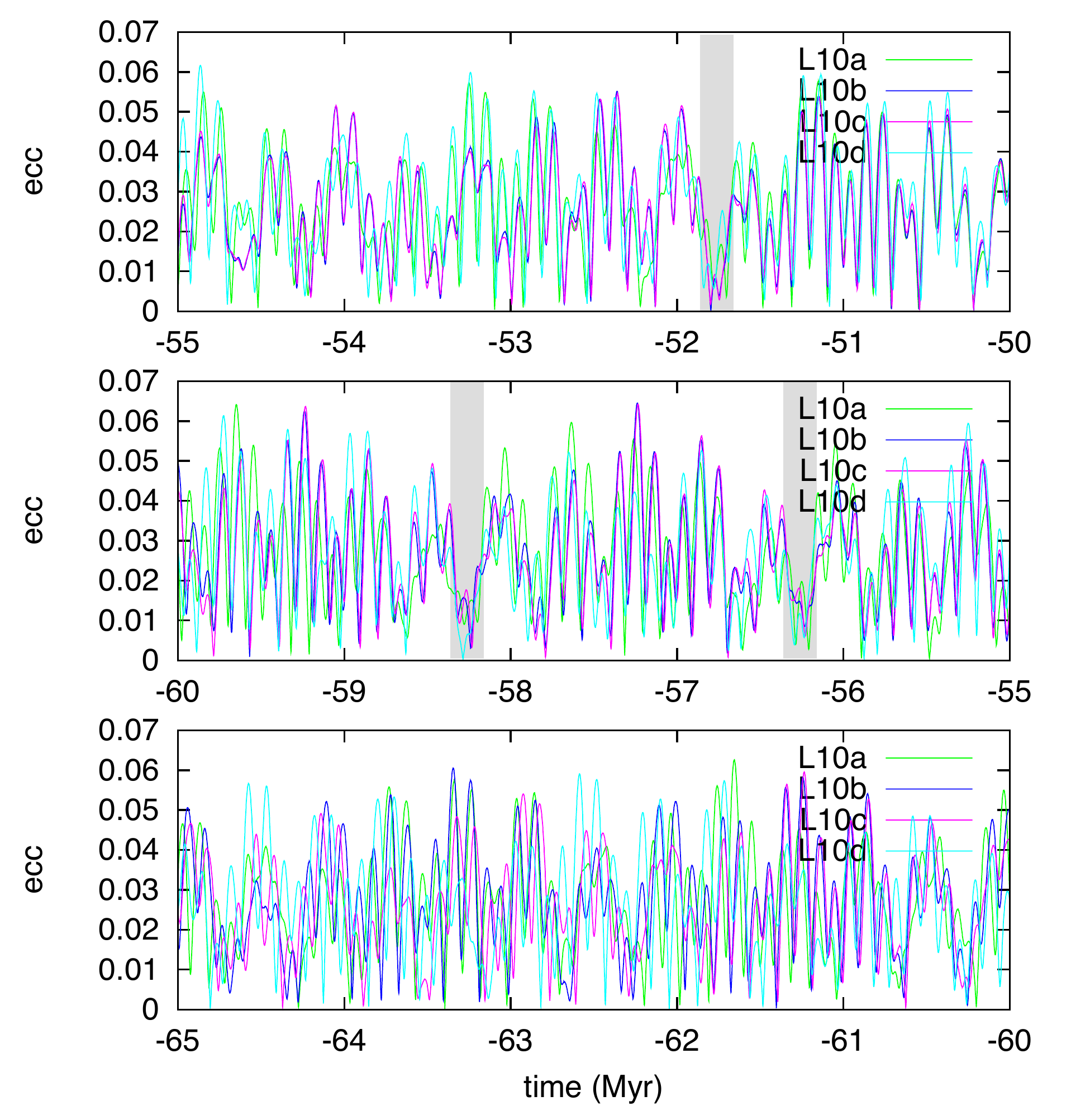} 
  \caption{
  Eccentricity of the Earth for the solutions La2010a (L10a),
  La2010b (L10b), La2010c (L10c), and La2010d (L10d) from $-50$ to $-65$ Myr.
  Although the various solution begin to diverge beyond 53 Ma, it is remarkable that the minima 
  of eccentricity at 51.75 Ma, 56.25 Ma and 58.25 Ma (in grey) correspond in all various solutions.
    } 
 \llabel{FigNc}
\end{figure*}
}

\newcommand\figNda{
\begin{figure*}[t]
 \includegraphics[width=15.cm]{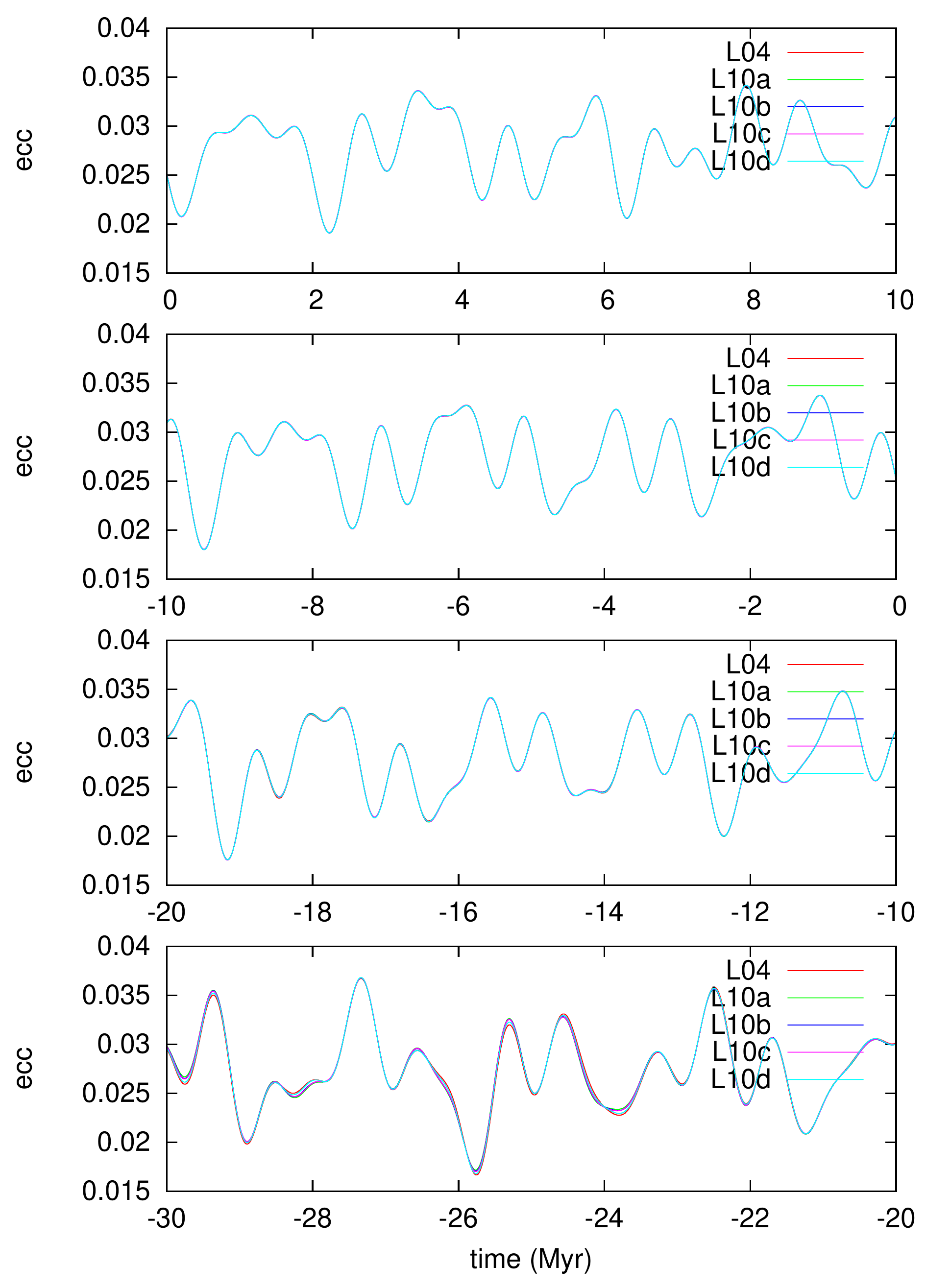} 
  \caption{
  Filtered eccentricity of the Earth for the solutions La2004 (L04), La2010a (L10a),
  La2010b (L10b), La2010c (L10c), and La2010d (L10d) from $+10$ to $-30$ Myr. 
  The solution is filtered in the interval $[0:2.2''/{\rm yr}]$, that is for 
  periods in $[589,+\infty[$ kyr.
    } 
 \llabel{FigNda}
\end{figure*}
}

\newcommand\figNd{
\begin{figure*}[t]
 \includegraphics[width=15.cm]{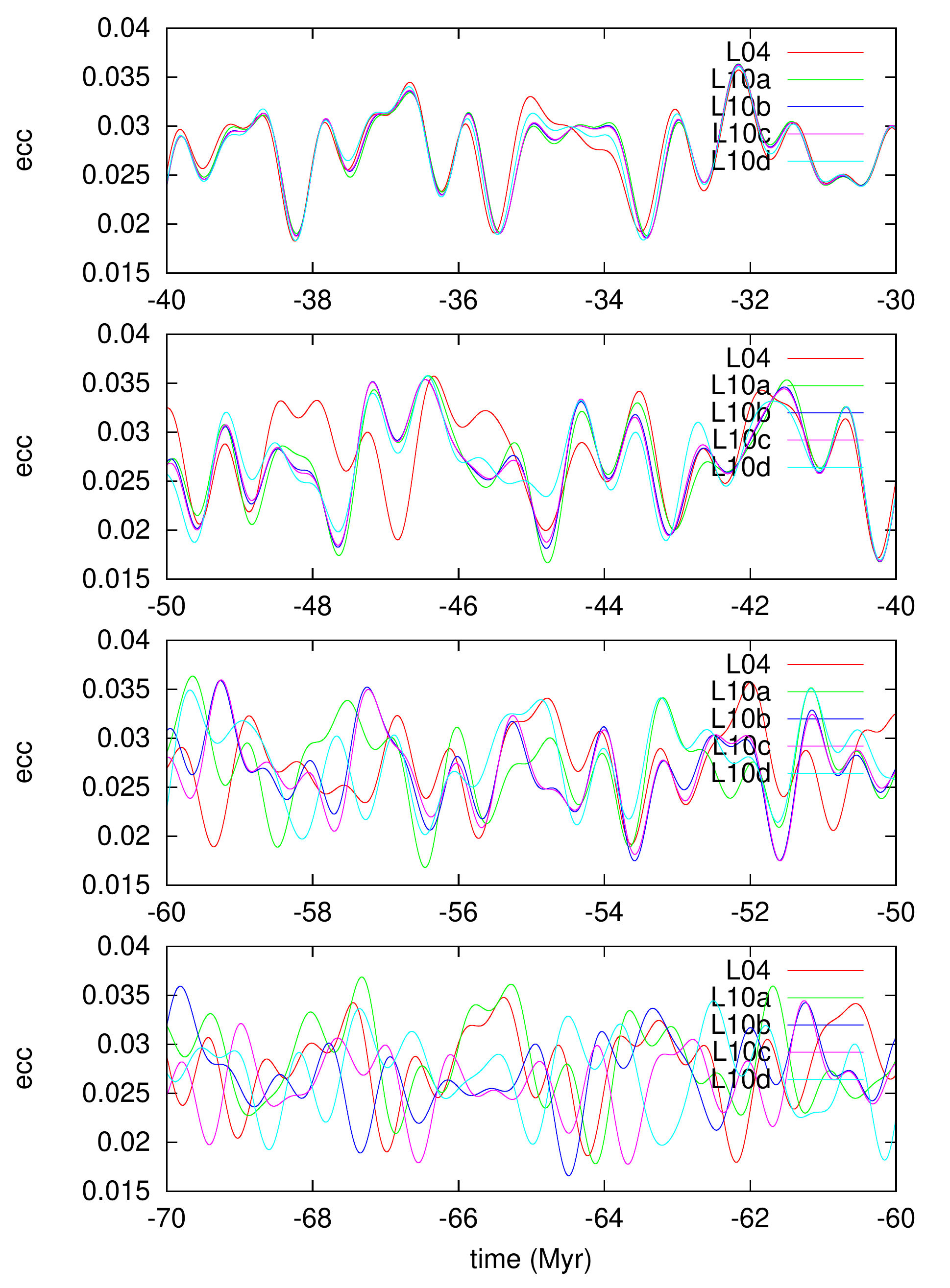} 
  \caption{
  Filtered eccentricity of the Earth for the solutions La2004 (L04), La2010a (L10a),
  La2010b (L10b), La2010c (L10c), and La2010d (L10d) from $-30$ to $-70$ Myr. 
  The solution is filtered in the interval $[0:2.2''/{\rm yr}]$, that is for 
  periods in $[589,+\infty[$ kyr.
    } 
 \llabel{FigNd}
\end{figure*}
}

\newcommand\figNe{
\begin{figure}[t]
 \includegraphics[width=8.cm]{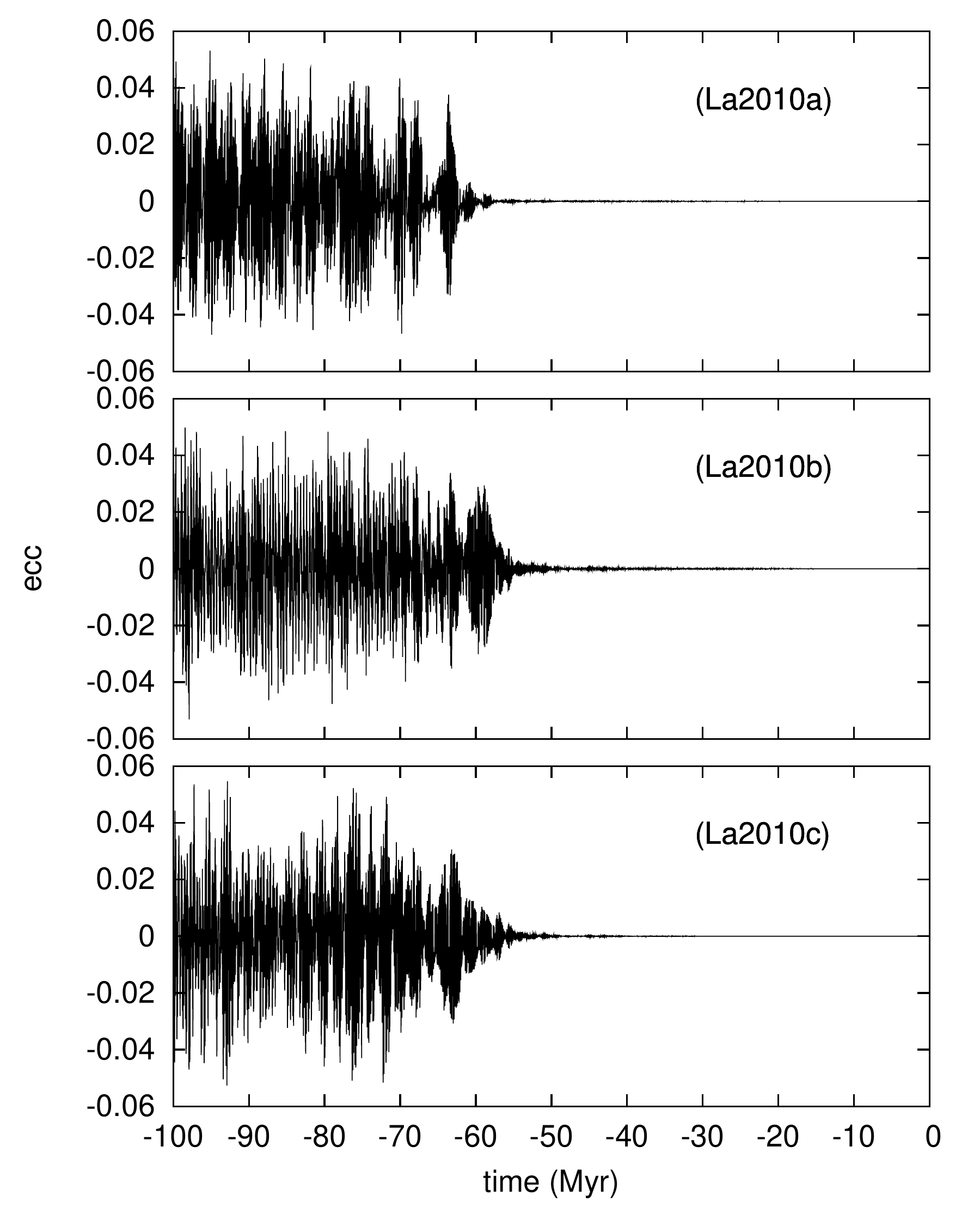} 
  \caption{
 Estimate of the numerical precision of the solutions La2010a,b,c.
 The estimate is obtained by the difference in the eccentricity of the Earth 
 obtained with the integration of two  solutions La2010x  and La2010x* 
 for x = a,b,c (see Tab. \ref{tab.solutions}). 
 } 
 \llabel{FigNe}
\end{figure}
}

\newcommand\figNf{
\begin{figure}[t]
 \includegraphics[width=8.cm]{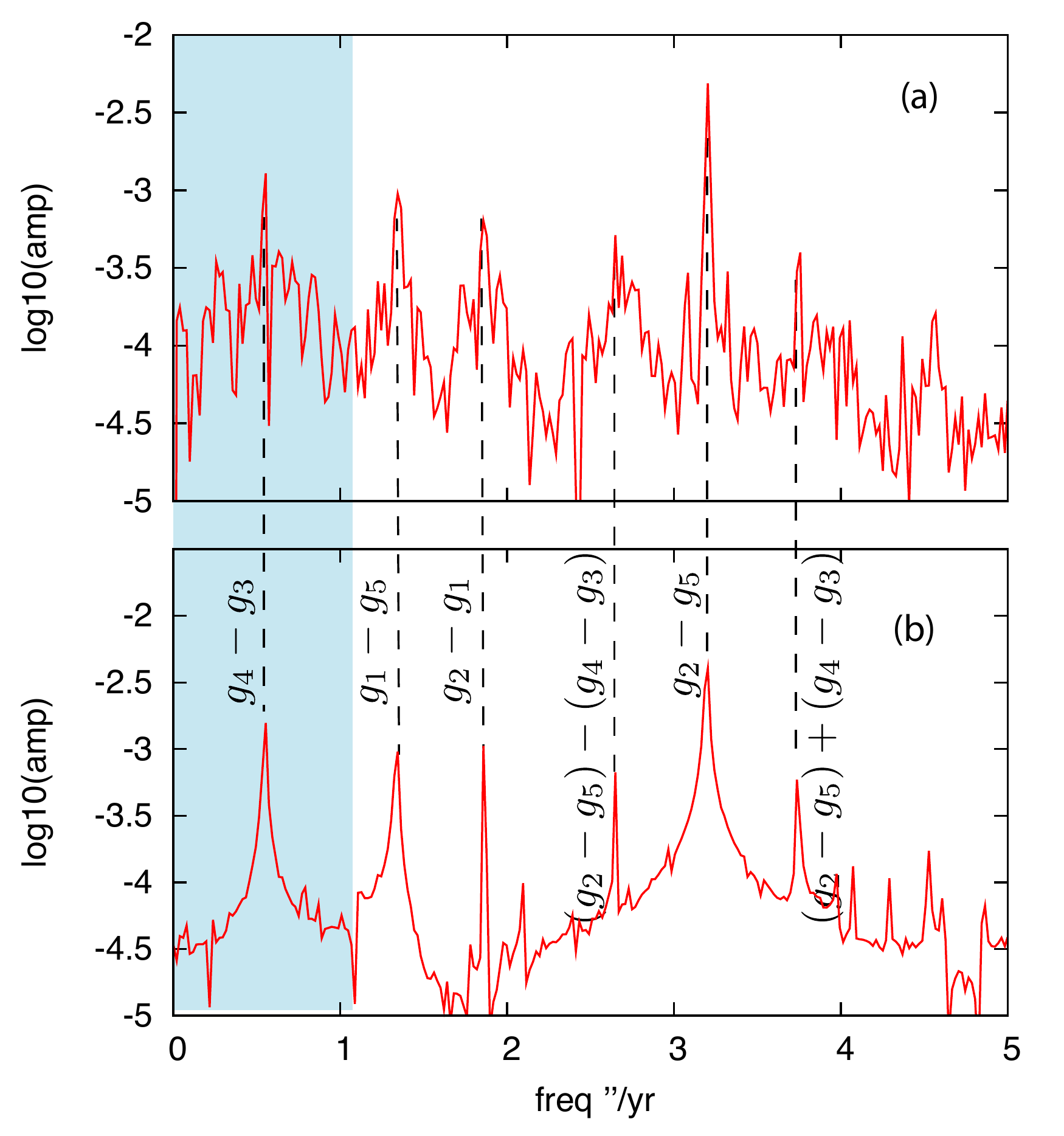} 
  \caption{
 In (a) is the spectrum of the eccentricity over 65Myr from the La2010a solution, limited to the interval 
 $[0,5''/yr]$. In (b), the same spectrum is plotted for a solution build with only the 5 main terms of 
 $z_3$ (Table \ref{tab.zecc}). The main peaks are then easily identified and thus also the main peaks of 
 the full eccentricity (a).} 
 \llabel{FigNf}
\end{figure}
}

\newcommand\figNg{
\begin{figure}[t]
 \includegraphics[width=8.cm]{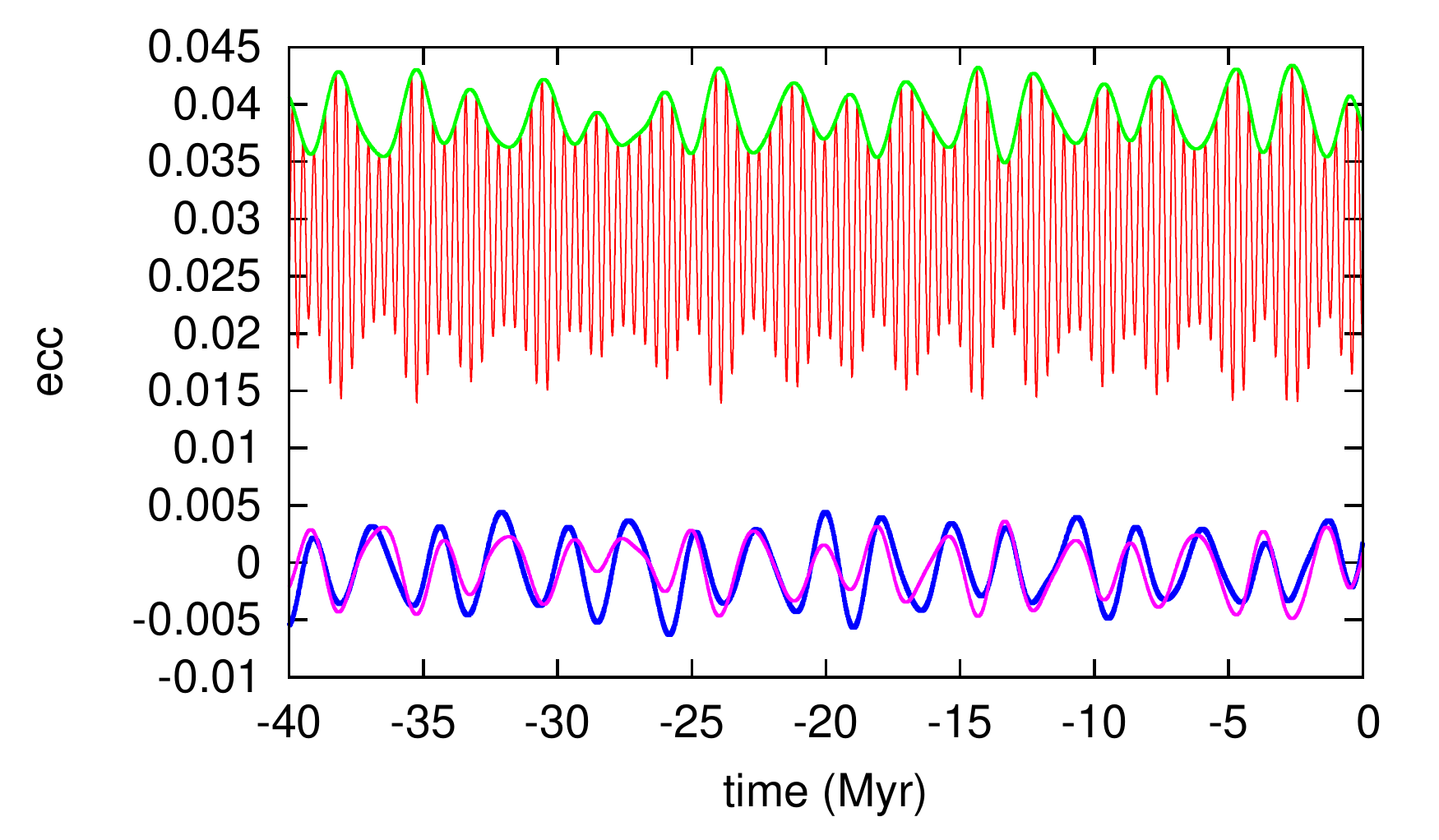} 
 \includegraphics[width=8.cm]{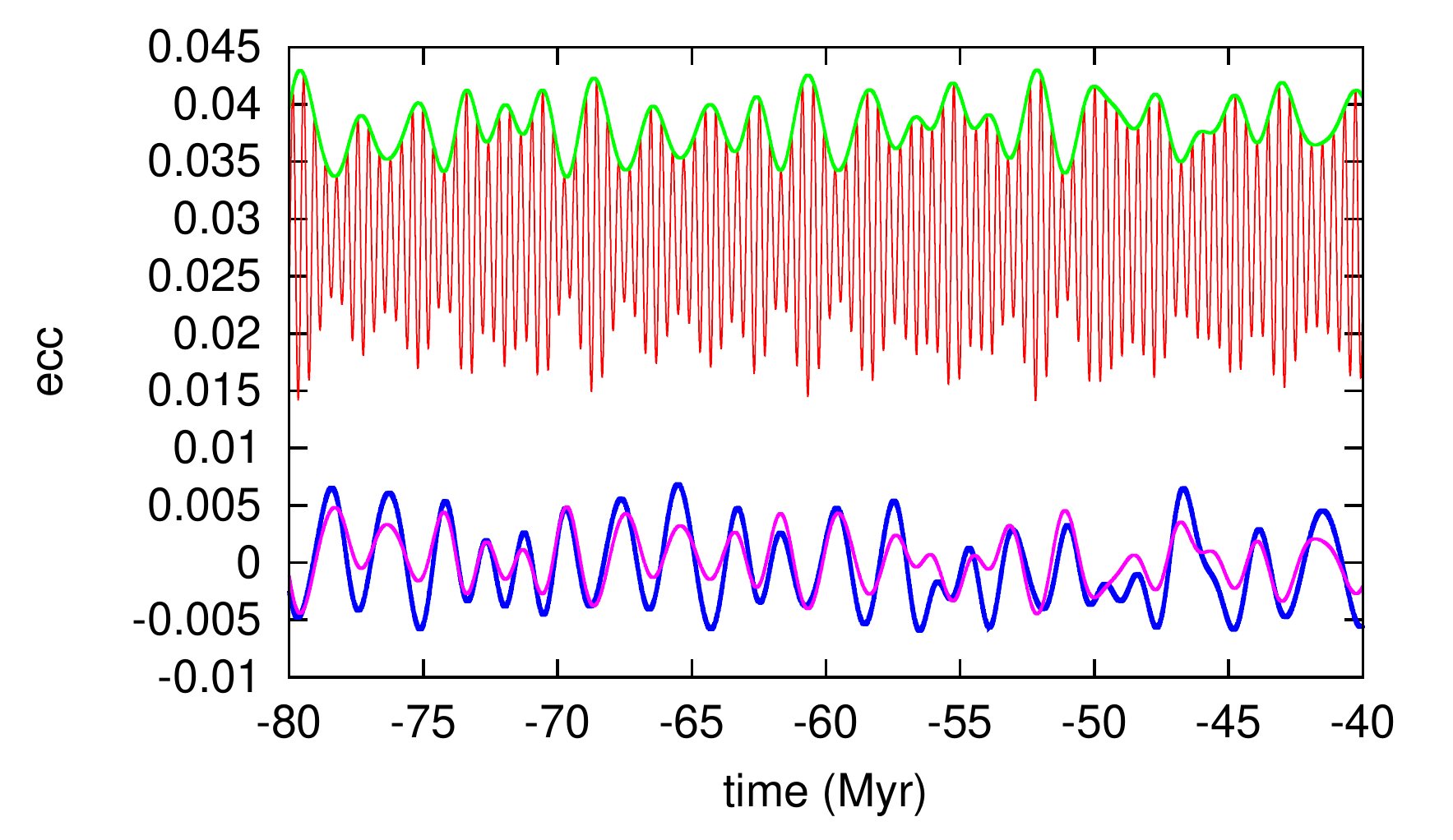} 
  \caption{
 Filtered eccentricity around the 405 kyr period for La2010a. In red is $e^a$, the filtered eccentricity 
 in the band $[2.2,4.3]$"/yr ($[301,589]$ kyr period), while $e^b$, the filtered eccentricity 
 in the band $[0,1.1]$ "/yr ( period $> 1.18$ Myr) is plotted in blue. 
 The opposite (in pink) of the maximum enveloppe of $e^a$ (in green)
 nearly coincide with $e^b$.} 
 \llabel{FigNg}
\end{figure}
}

\newcommand\figNgg{
\begin{figure}[t]
 \includegraphics[width=8.cm]{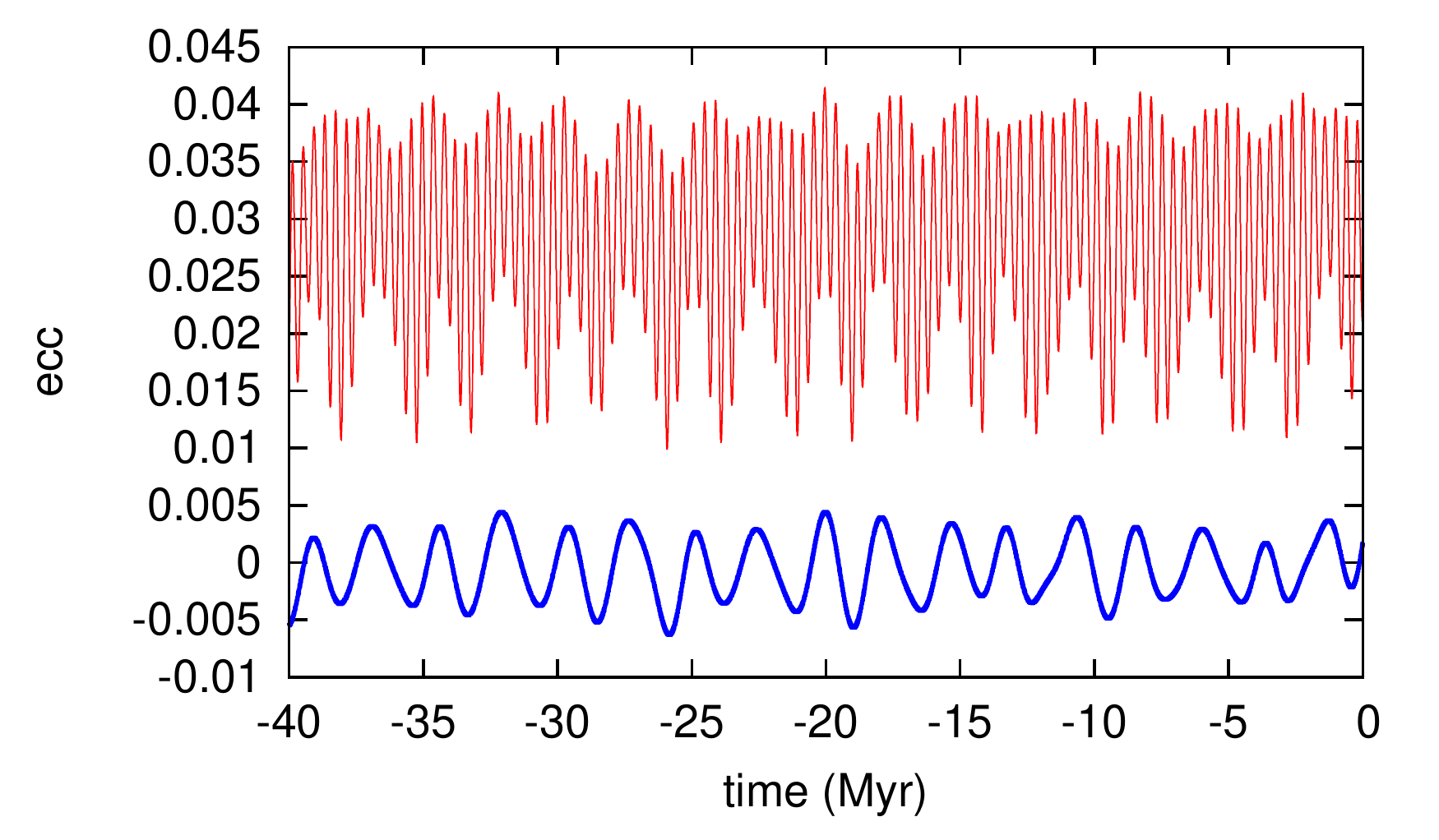} 
 \includegraphics[width=8.cm]{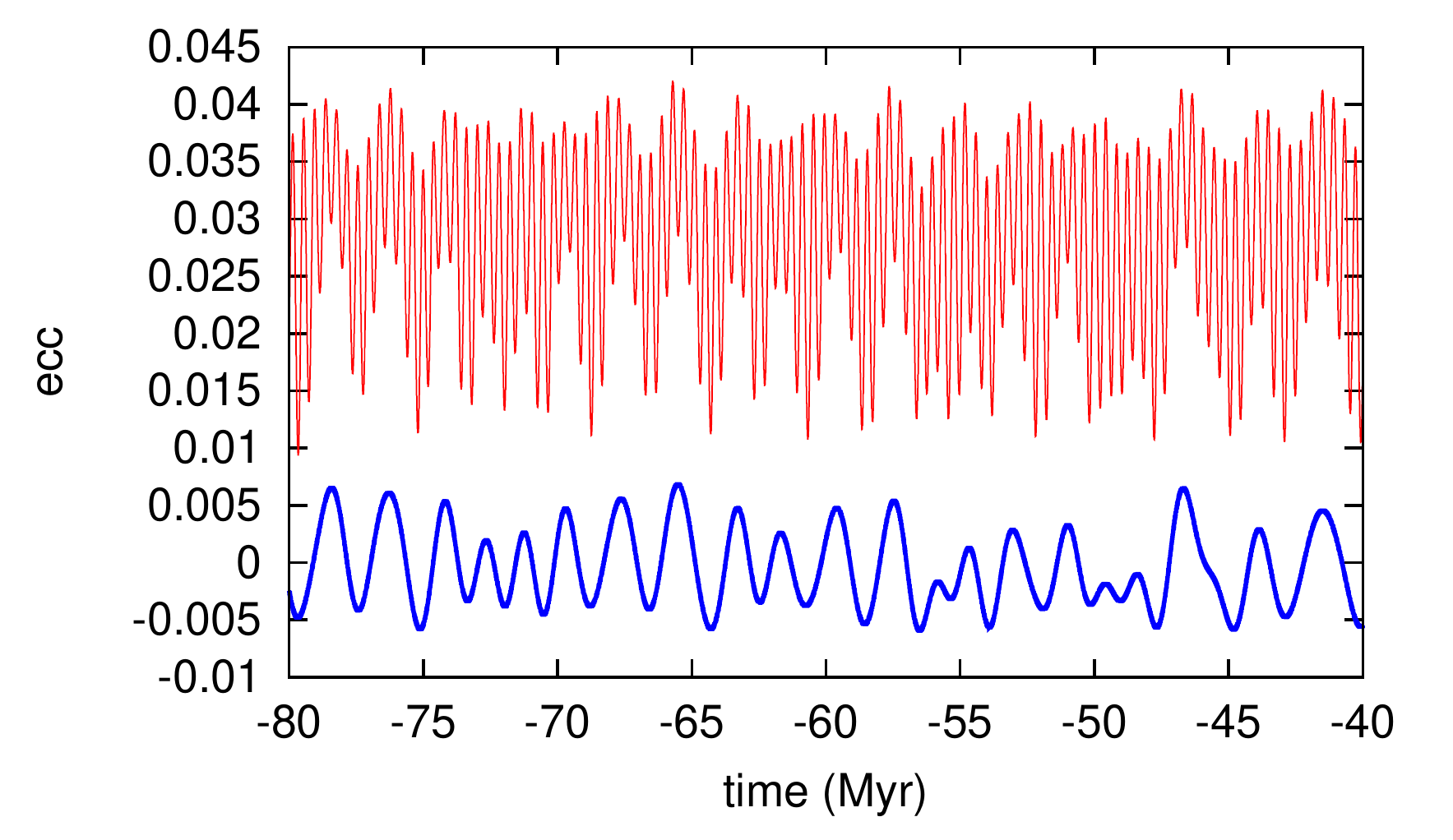} 
  \caption{
  $e^a + e^b$ (see Fig. \ref{FigNg}) for the La2010a solution is plotted in red. Its minimum 
  are in phase with  $e^b$, the filtered eccentricity 
 in the band $[0,1.1]$ "/yr (in blue). 
 } 
 \llabel{FigNgg}
\end{figure}
}

\newcommand\figNh{
\begin{figure}[t]
 \includegraphics[width=8.3cm]{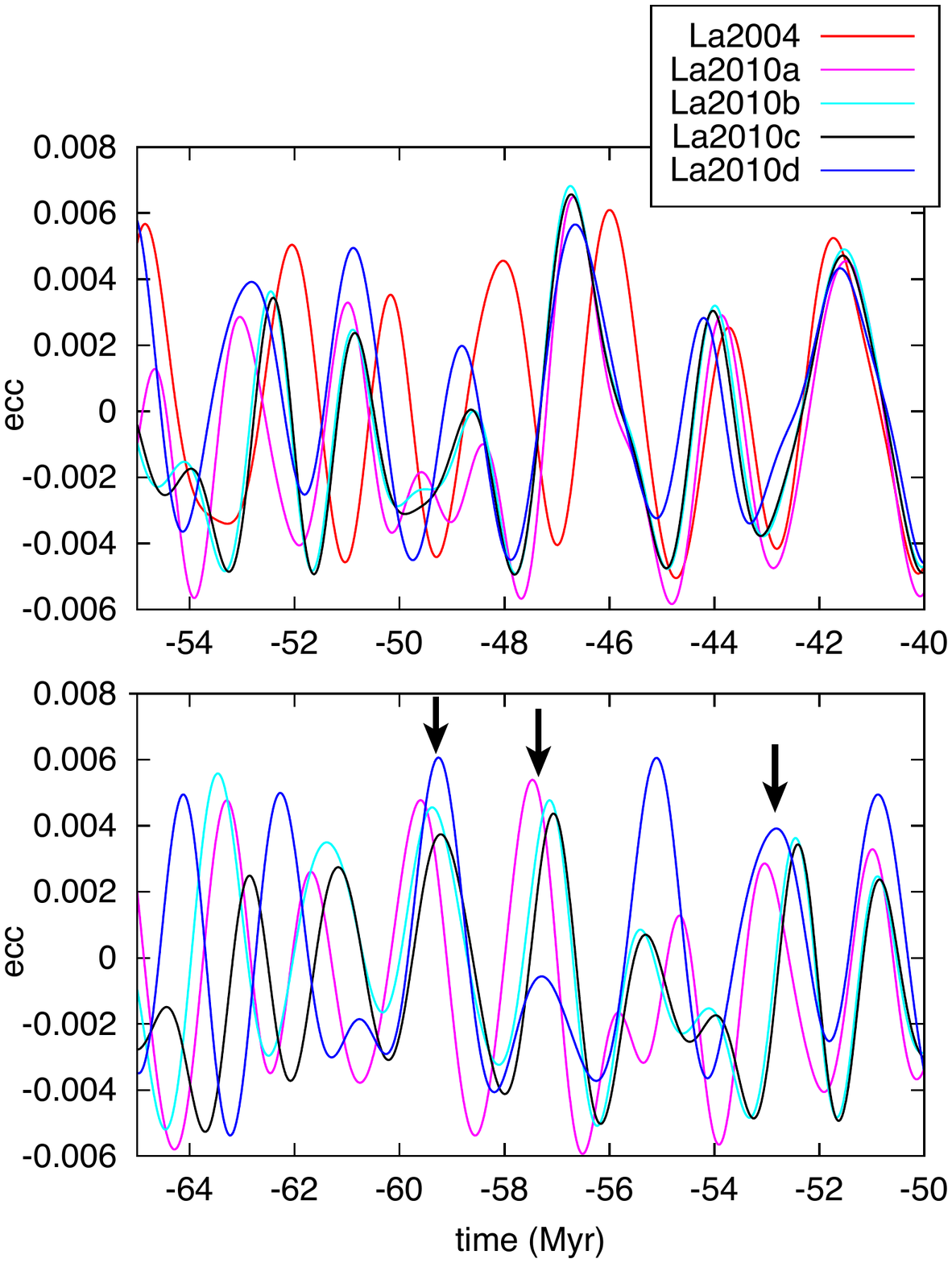} 
  \caption{
 Filtered eccentricity 
 in the band $[0,1.1]$ "/yr ( period $> 1.18$ Myr)  for La2004 and  La2010a,b,c,d.} 
 \llabel{FigNh}
\end{figure}
}

\newcommand\figNi{
\begin{figure}[t]
 \includegraphics[width=8.3cm]{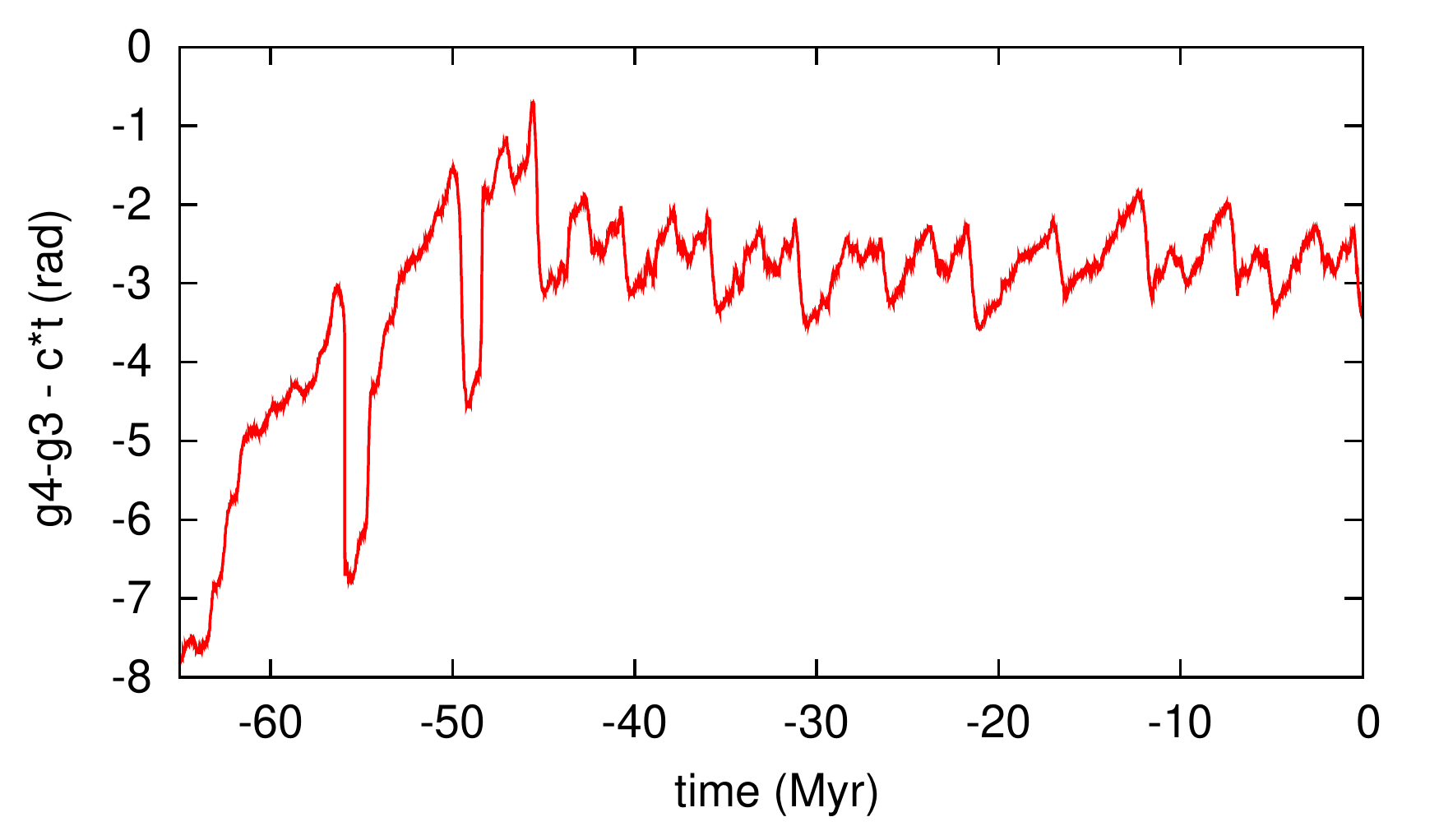} 
 \includegraphics[width=8.3cm]{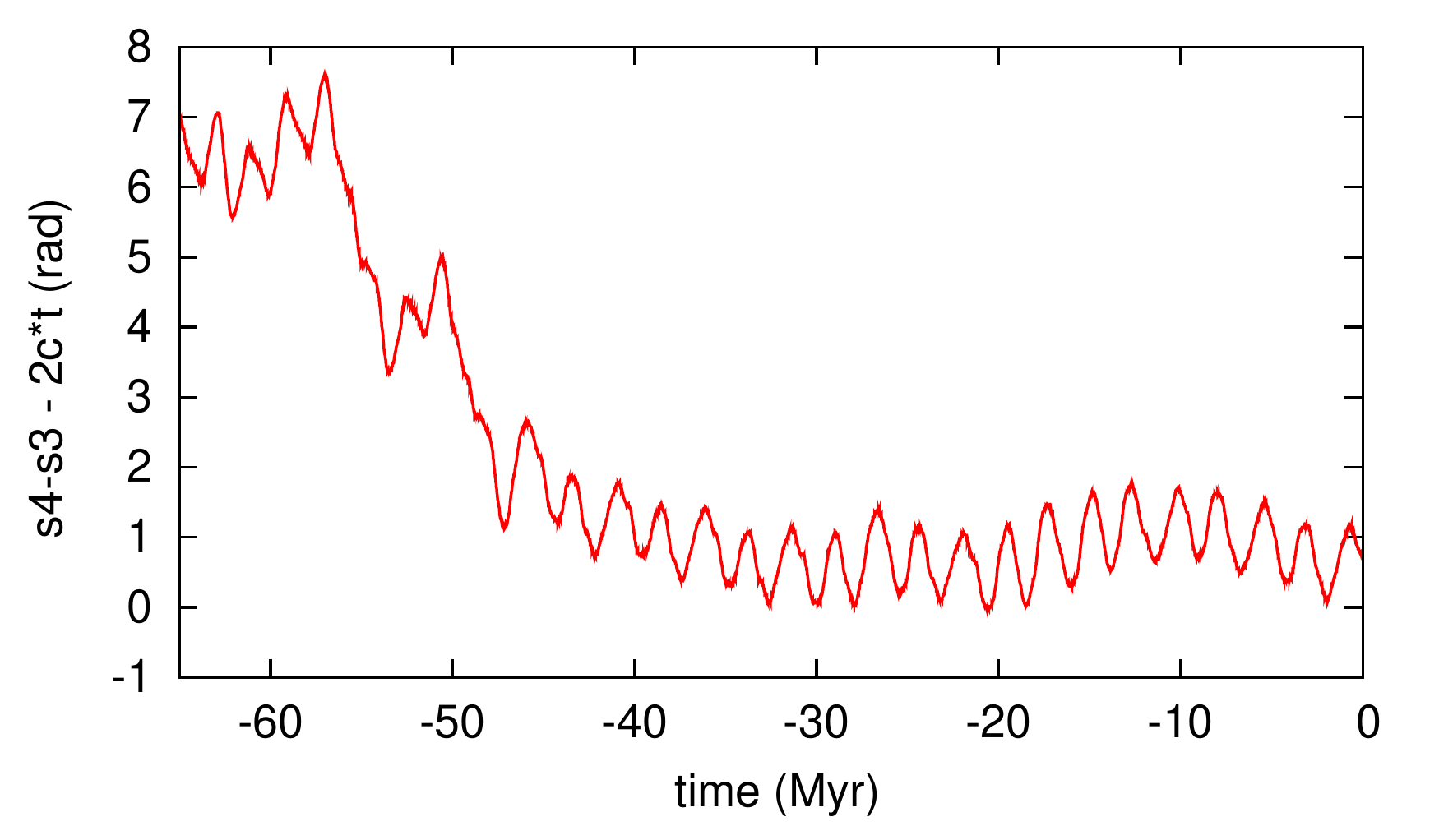} 
  \caption{
Argument (in radians) related to 
 $g_4-g_3 - 2.664 \, T $ versus $T$ (in Myr) for La2010a (top).
Argument (in radians) related to 
 $s_4-s_3 - 2\times 2.664 \, T $ versus $T$ (in Myr) for La2010a (bottom).
 }
 \llabel{FigNi}
\end{figure}
}

\newcommand\figNj{
\begin{figure}[t]
 \includegraphics[width=8.3cm]{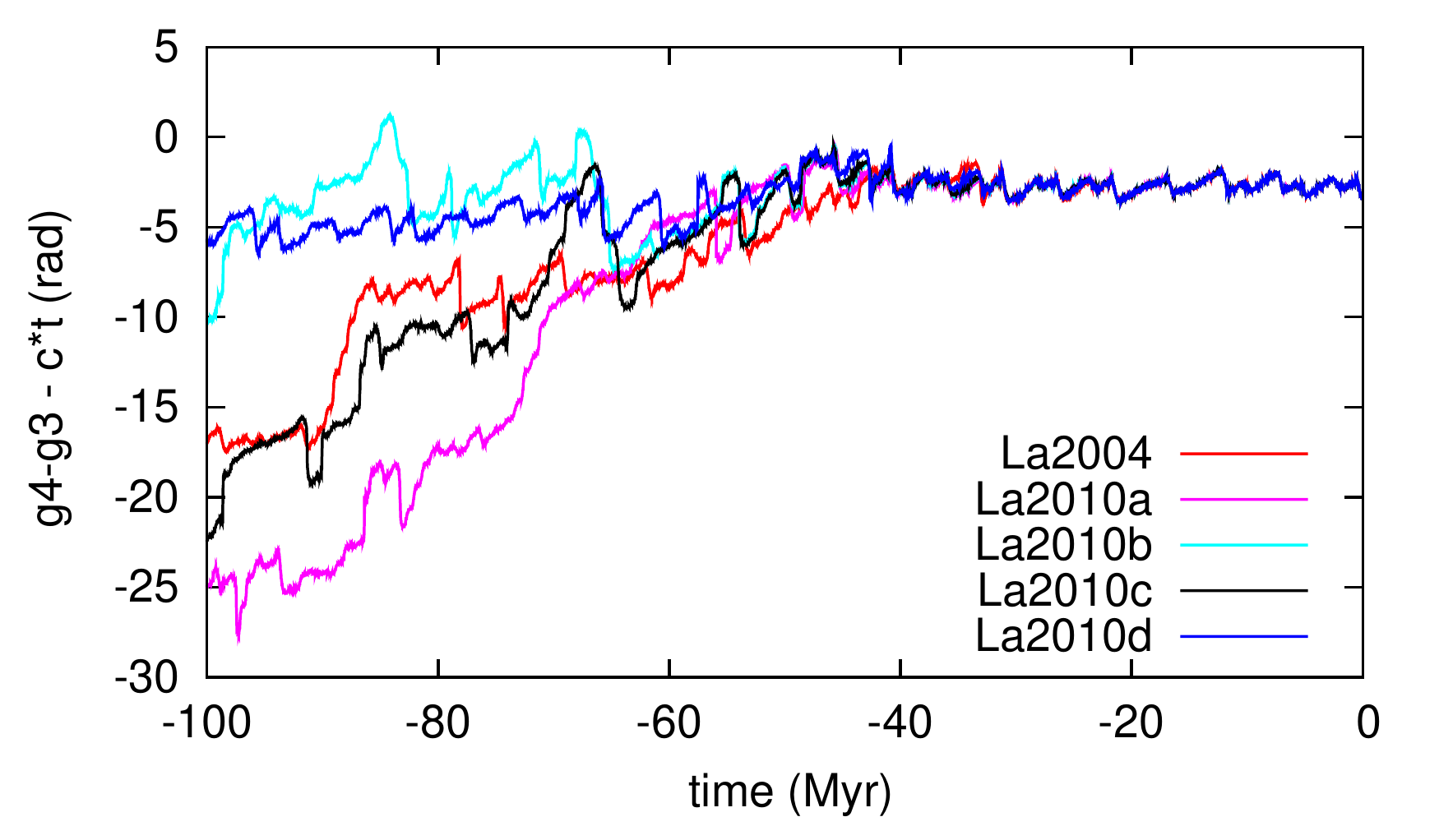} 
 \includegraphics[width=8.3cm]{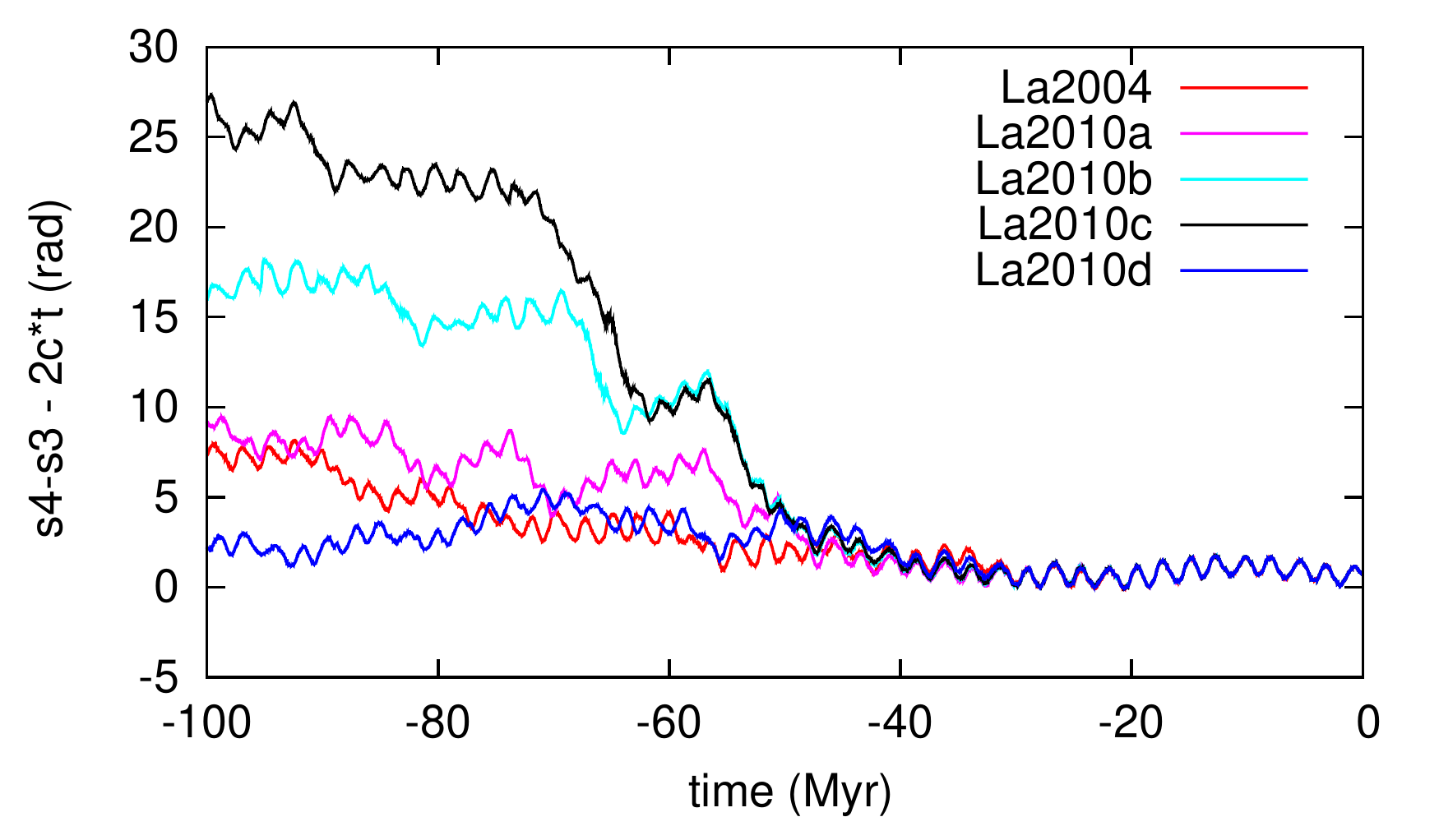} 
  \caption{
Argument (in radians) related to 
 $g_4-g_3 - 2.664 \, T $ versus $T$ (in Myr) for various orbital solutions (top).
Argument (in radians) related to 
 $s_4-s_3 - 2\times 2.664 \, T $ versus $T$ (in Myr) for various orbital solutions (bottom).
  } 
 \llabel{FigNj}
\end{figure}
}

\newcommand\tabNa{
\begin{table}
\caption{Maximum difference between INPOP06 and INPOP08  over 1 Myr (top).
and between INPOP08 (computed in extended precision) and INPOP08d, a version of 
INPOP08 computed in double precision, over 1 Myr (bottom).}
\def\espace{\rule[-5pt]{0pt}{16 pt} }
\begin{tabular}{|r|rrr|}
\hline
       & \multicolumn{3}{c|}{INPOP06-INPOP08   }  \\
\hline  
        &     $a \times 10^{6}$ &  $e \times 10^{6} $ & $i \times 10^{6}$ \\
\hline
Mercury &       0.027  &       25.544  &      12.598 \\ 
Venus   &       0.543  &        4.746  &       6.299 \\ 
EMB     &       1.067  &        4.709  &       4.029 \\ 
Mars    &       3.852  &        7.134  &       2.570 \\ 
Jupiter &      56.126  &       20.542  &       0.577 \\ 
Saturn  &     585.092  &       76.138  &       1.315 \\ 
Uranus  &     885.497  &       92.313  &       1.481 \\ 
Neptune &    3449.727  &      104.593  &       0.882 \\ 
Pluto   &   19900.821  &      297.483  &      46.579 \\ 
Moon    &      57.345  &    42006.038  &    1463.465 \\ 
\hline
&\multicolumn{3}{c|}{INPOP08-INPOP08d }\\
\hline
        &     $a \times 10^{6}$ &  $e \times 10^{6} $ & $i \times 10^{6}$ \\
\hline
Mercury &        0.012  &        0.077  &       0.011 \\ 
Venus   &        0.640  &        1.706  &       0.138 \\ 
EMB     &        0.889  &        1.768  &       0.125 \\ 
Mars    &        6.899  &        6.081  &       0.394 \\ 
Jupiter &        1.004  &        0.175  &       0.009 \\ 
Saturn  &        2.542  &        0.243  &       0.007 \\ 
Uranus  &        6.834  &        0.333  &       0.006 \\ 
Neptune &       13.479  &        0.415  &       0.008 \\ 
Pluto   &       25.332  &        0.455  &       0.061 \\ 
Moon    &       29.594  &    12400.799  &     527.701 \\ 
\hline
\end{tabular}
\llabel{tab.inpop}

{{\bf Note.} EMB is the Earth-Moon barycenter. The semi-major axis $a$ is in AU  and the inclination
with respect to the invariable plane 
$i$ is in radians. }
\end{table}
}

\newcommand\tabNb{
\begin{table}
\def\espace{\rule[-5pt]{0pt}{16 pt} }
\begin{tabular}{|r|rrr|}
\hline
       & \multicolumn{3}{c|}{ast5AL08cx -INPOP08   1 Ma  }  \\
\hline  
        &     $a \times 10^{6}$ &  $e \times 10^{6} $ & $i \times 10^{6}$ \\
\hline
Mercury &        0.006  &        0.583  &       0.498 \\ 
Venus   &        0.341  &        1.027  &       0.351 \\ 
EMB     &        0.453  &        1.182  &       0.343 \\ 
Mars    &        1.774  &        1.608  &       0.351 \\ 
Jupiter &        0.551  &        0.147  &       0.049 \\ 
Saturn  &        1.758  &        0.359  &       0.037 \\ 
Uranus  &        5.129  &        0.478  &       0.029 \\ 
Neptune &        8.734  &        0.306  &       0.027 \\ 
Pluto   &       17.901  &        0.325  &       0.049 \\ 
Moon    &        8.916  &    22005.375  &   12538.037 \\ 
\hline
&\multicolumn{3}{c|}{ast5SL08ax-INPOP08    1 Ma}\\
\hline
        &     $a \times 10^{6}$ &  $e \times 10^{6} $ & $i \times 10^{6}$ \\
\hline
Mercury &        0.002  &        3.962  &       2.859 \\
Venus   &        0.155  &        1.975  &       0.978 \\
EMB     &        0.224  &        1.540  &       0.968 \\
Mars    &        1.744  &        2.823  &       0.954 \\
Jupiter &        0.278  &        0.113  &       0.033 \\
Saturn  &        0.777  &        0.273  &       0.023 \\
Uranus  &        2.256  &        0.352  &       0.020 \\
Neptune &        4.052  &        0.147  &       0.009 \\
Pluto   &        8.805  &        0.165  &       0.019 \\
\hline
\end{tabular}
\caption{Maximum difference between the La2010 solutions  and INPOP08  over 1 Myr.
Top:  Maximum difference  ast5AL08cx -INPOP08   over 1 Myr. 
Bottom : Maximum difference  ast5SL08ax -INPOP08   over 1 Myr. In ast5SL08ax, the Moon contribution is 
averaged. All solutions are in extended precision.
EMB is the Earth-Moon barycenter. The semi-major axis $a$ is in AU  and the inclination with respect 
to the invariable plane 
$i$ is in radians. }
\llabel{tab.Nb}
\end{table}
}

\newcommand\tabNc{
\begin{table}
\def\espace{\rule[-5pt]{0pt}{16 pt} }
\begin{tabular}{|r|rrr|}
\hline
       & \multicolumn{3}{c|}{ast5ALix -INPOP06   1 Ma  }  \\
\hline  
        &     $a \times 10^{6}$ &  $e \times 10^{6} $ & $i \times 10^{6}$ \\
\hline
Mercury &         0.005  &        0.473  &       0.342 \\
Venus   &         0.034  &        0.503  &       0.209 \\
EMB     &         0.049  &        0.505  &       0.234 \\
Mars    &         0.721  &        0.873  &       0.397 \\
Jupiter &         0.121  &        0.069  &       0.046 \\
Saturn  &         0.980  &        0.157  &       0.032 \\
Uranus  &         1.553  &        0.139  &       0.025 \\
Neptune &         2.263  &        0.068  &       0.020 \\
Pluto   &         7.068  &        0.104  &       0.022 \\
Moon    &        21.439  &    26686.705  &   12769.784 \\
\hline
&\multicolumn{3}{c|}{ast5ALh-INPOP08    1 Ma}\\
\hline
        &     $a \times 10^{6}$ &  $e \times 10^{6} $ & $i \times 10^{6}$ \\
\hline
Mercury &        0.007  &        0.503  &       0.357 \\
Venus   &        0.290  &        1.156  &       0.225 \\
EMB     &        0.379  &        0.986  &       0.257 \\
Mars    &        2.757  &        2.484  &       0.469 \\
Jupiter &        0.563  &        0.139  &       0.047 \\
Saturn  &        2.406  &        0.272  &       0.034 \\
Uranus  &        5.415  &        0.268  &       0.026 \\
Neptune &        9.845  &        0.320  &       0.021 \\
Pluto   &       19.768  &        0.361  &       0.052 \\
Moon    &       21.093  &    26481.398  &   12727.162 \\
\hline
\end{tabular}
\caption{Maximum difference between the La2010 solutions  and INPOP06  over 1 Myr.
Top: Maximum difference  ast5ALix -INPOP06 (Both solutions are in extended precision) over 1 Myr.
Bottom: Maximum difference  ast5ALh -INPOP06 over 1 Myr. The solution ast5ALh has been computed in double precision, while 
INPOP06 is in extended precision. Both solutions have been fitted to INPOP06 over 1 Myr.
EMB is the Earth-Moon barycenter. The semi-major axis $a$ is in AU  and the inclination with respect to the invaraible plane 
$i$ is in radians. }
\llabel{tab.Nc}
\end{table}
}

\newcommand\tabNi{
\begin{table*}
\caption{Variants of the La2010 solutions. The nominal solution La2010a
is obtained for a stepsize $\tau=1 \times 10^{-3}$ in extended precision. The other solutions, with 
different setting have been computed to test 
the stability  of the nominal solution.}
\begin{tabular}{|l|l|c|l|l|l|l|}
\hline
name      &   files       & ephem    & fit        &   $ \tau$    (yr)       &  prec  &  \\ 
\hline 
La2010a   & ast5AL08cxc  & INPOP08a  & 0.58 Myr    & $1 \times 10^{-3}$          & Ext & 5ast,M \\
La2010a*  & ast5AL08czc  & INPOP08a  & 0.58 Myr    & $0.9765625 \times 10^{-3}$  & Ext & 5ast,M \\
La2010b   & ast5AL08cx   & INPOP08a  & 0.58 Myr    & $5 \times 10^{-3}$          & Ext & 5ast,M \\
La2010b*  & ast5AL08cz   & INPOP08a  & 0.58 Myr    & $4.8828125 \times 10^{-3}$  & Ext & 5ast,M \\
La2010c   & ast0AL08cx2a & INPOP08a  & 1 Myr    & $5 \times 10^{-3}$          & Ext & 0ast,M \\
La2010c*  & ast0AL08cz2a & INPOP08a  & 1 Myr    & $4.8828125 \times 10^{-3}$  & Ext & 0ast,M \\
La2010d   & ast5ALix     & INPOP06  & 1 Myr       & $5 \times 10^{-3}$          & Ext & 5ast,M \\
\hline
\end{tabular}
\label{tab.solutions}

{\bf Note.} The column "files" denotes the name of the computer files of the solution. "ephem" is the 
name of the reference INPOP ephemeris. "fit" is the time length of the ephemeris used for the fit. 
$\tau$ is the step size of the numerical integration. In "prec", Ext indicates that the integration 
was performed in extended  precision. In the last column, 5ast indicates that 5 asteroids have been integrated,
and M stands for the Moon as a separate object. 
\end{table*}
}

\title{La2010: A new orbital solution for the long term motion of the Earth.}
\author{J. Laskar\inst{1}
\and A. Fienga\inst{1,2}
\and M. Gastineau\inst{1}
\and H. Manche\inst{1}
 }

\institute{ ASD, IMCCE-CNRS UMR8028, Observatoire de Paris, UPMC, 
77 Av. Denfert-Rochereau, 75014 Paris, France
\and
 Observatoire de Besan\c con--CNRS UMR6213,
 41bis Av. de l'Observatoire, 25000 Besan\c con
}
\offprints{J. Laskar, \email{laskar@imcce.fr}}

\titlerunning{Orbital solution of the Earth}
\authorrunning{Laskar \etal}

\date{\today}
\abstract{
We present here a new solution for the astronomical computation of the orbital motion 
of the  Earth spanning from 0 to $-250$ Myr.
The main improvement  with respect to  the previous numerical solution La2004 \citep{LaskRobu2004a} 
is an improved adjustment of the parameters and initial conditions through a 
fit  over 1 Myr  to a special version of the high accurate numerical ephemeris INPOP08 
\citep{FienLask2009a}. The precession equations have also been entirely revised and 
are no longer averaged over the orbital motion of the Earth and Moon. 
This new orbital solution is now valid over more than 50 Myr in the past or in the future 
with proper phases of the eccentricity variations. 
Due to chaotic behavior, the precision of the solution decreases rapidly beyond 
this time span, and we discuss  the behavior of various solutions beyond 50 Myr. 
For paleoclimate calibrations, we provide several different solutions that are all compatible with the most 
precise planetary ephemeris. We have thus reached the time where geological data are now required 
to discriminate among planetary orbital solutions beyond 50 Myr.  
 }
  
\begin{document}
\maketitle
%\tableofcontents
\section{Introduction}

Due to  gravitational planetary perturbations, the elliptical elements of
the orbit of the Earth are slowly changing in time, as is the orientation of 
the planet's spin axis. As described by 
\citep{Mila1941a} these changes induce variations of the insolation received 
on the Earth's surface that are at the origin of large climatic changes. 
Since the work of \citep{HaysImbr1976a}, that established a correlation 
between   astronomical forcing and the $\delta^{18}O$ records over the
past 500 kyr, there has been a increasing need for precise long term 
ephemeris for the Earth orbital and rotational evolution (see \citet{LaskRobu2004a}
for a more detailed historical account). 

 For paleoclimate studies, the most widely used orbital solutions are nowadays either the 
averaged solution of \citep{Lask1988a,LaskJout1993a} or the most recent numerical solution 
of \citet{LaskRobu2004a}.

The first long term direct numerical integration (without averaging) 
of a realistic model of the Solar system, together with the 
precession and obliquity equations, 
was made by \citep{QuinTrem1991a}  over 3 Myr.   
Over its range, this solution presented   small differences with 
the  secular solution of \citep{Lask1988a,Lask1990a}, (see \citet{LaskQuin1992a}). 
The orbital motion of the full Solar system has 
then been computed over 100 Myr by \citep{SussWisd1992a}, using 
a symplectic integrator with mixed variables \citep{WisdHolm1991a},
 confirming the chaotic behavior found by \citet{Lask1989a,Lask1990a}.
Following  the improvement of computer technology, long term integrations 
 of realistic models of the Solar system have improved \citep{VaraRunn2003a,LaskRobu2004a}, 
 but the main limitation remains the exponential divergence of 
 nearby orbits resulting from the chaotic motion of the Solar system 
 \citep{Lask1989a,Lask1990a,Lask1999a}.
  Although it is now possible to integrate the motion of the Solar system over time 
 periods of more than 5 Gyr, comparable to its age or expected life time 
\citep{LaskGast2009a}, it is clear that  the chaotic behavior of 
 the solution will still limit its validity to a few tens of Myr.

 The present paper is a continuation of the work that has been conducted for decades in our group 
 in order to obtain the most precise solution for  the past evolution of the orbit and rotational 
 state of the Earth, aimed to paleoclimate studies. 
 
 The numerical integrator is the same symplectic integrator from \citep{LaskRobu2001a} 
 as the one used in the La2004 solution \citep{LaskRobu2004a}, but it was entirely 
 rewritten in C in order to access to extended precision on the intel architecture.
 On the other hand, the tidal model has been largely modified, and is now close to the one 
 used in  the JPL planetary ephemeris DE405 \citep{Stan1998a} or in our new planetary ephemeris 
 INPOP \citep{FienManc2008a,FienLask2009a}. 
 The precession equations for the evolution of the spin axis of the Earth are also new
 \citep{BoueLask2006a}. There are no longer averaged over the orbital motion of the 
 planets, and allow a precise computation of the evolution of the Earth spin axis
 that can be compared to the most precise model adopted by the IAU  \citep{SoffKlio2003a}
 \citep[see][]{FienManc2008a}. 
 
 In previous long term solutions \cite{Lask1988a,QuinTrem1991a,LaskJout1993b,VaraRunn2003a,LaskRobu2004a}, 
 the initial conditions of the solutions were obtained 
 either directly from a high precision planet ephemeris, or by a fit over its 
 full time span (as in La2004) that was still limited to a few thousands of years. 
It was thus difficult to monitor the real uncertainty of the used ephemeris. 
 
 In the present work, we have profoundly removed this limitation. Indeed, 
 in the past few years, we have entirely build a new high precision planetary 
 ephemeris INPOP that has been fitted to all available planetary and Lunar 
 observations \citep{FienManc2008a,FienLask2009a}.  This ephemeris that has been released already 
 in two versions (INPOP06 and INPOP08) is thus equivalent to the JPL ephemerides 
 DE that were used for determining the initial conditions of the previous 
 long term solutions. In addition, we have removed in INPOP all time limitations 
 and carefully designed the numerical integrator. We could thus extend the integration of INPOP06 and INPO08 
 over 1 Myr with the full ephemeris model. 
 The  initial conditions of the present long term ephemeris could then be fitted over 
 an extended interval of several hundreds of thousands of years before being
 extended to 250 Myr. By doing so, we were able to  take into account the 
 full precision of the ephemeris, and it appears now that the 
 limitations is no longer in the model but in the planetary observations themselves. 
 
 The first sections of this paper (sec 2 to sec 6 ) describe successively the La2010 numerical model,
 its link with the INPOP ephemeris, the various La2010 solutions and their comparison with 
 the high precision INPOP ephemeris and the previous La2004 solution. The following sections (7 to 9) 
 are focussed on the long term cycles that are present in the eccentricity solution and their 
 stability. This topic is of essential importance for the attempt to establish an astronomically 
 calibrated geological timescale \citep[see][]{paelike_rock_2008} . Indeed, the La2004 solution \citep{LaskRobu2004a} has been 
 successfully used for the astronomical calibration of the Neogene period ($\approx 23$ Myr)
 \citep{L.J.Wils2004a} that is included in the most recent standard timescale GTS2004  
 \citep{GradOgg2004a}.  At present, there is a continuous effort to improve 
 this timescale and to extend the astronomical calibration to the full 
 Cenozoic period ($\approx 65$ Myr) through the Earthtime and Earthtime-eu projects 
 (http://www.earth-time.org, http://earthtime-eu.eu). In order  to do so, the length of validity 
 of the orbital solution has to be extended by more than 20 Myr. Because of the chaotic behavior of the 
 solution  \citep{Lask1989a},  this  corresponds to improve the precision of the model and parameters 
 by two orders of magnitude, and the present work is an attempt in this direction. 
 
 On the other hand, beyond the horizon of predictibility of the orbital solution, it is tempting to
 use the recorded geological information to provide constraints on the orbital motion 
 of the Solar System \citep{LourWeha2001a,PaliLask2004a}. Due to the lack of precision of the geological data in remote periods of time, 
 this can only be done  through macroscopic aspects of the orbital solution. 
The analysis of the secular resonance $g_4-g_3 -2(s_4-s_3)$ (Sec. 8) is devoted to this problem. 
In particular, it is shown how the analysis of the modulation of the amplitude of the  405 kyr 
eccentricity term can discriminate among various orbital solutions and thus provide feedback 
from geological data  to astronomical models.

\section{Numerical model}
\llabel{sec.A}

The orbital solutions La90-93 \citep{Lask1990a,LaskJout1993b} were obtained 
by a numerical integration of the averaged equations of the Solar system, 
including  the main general relativity and Lunar perturbations. As 
computer technology now allows to integrate directly  precise models 
of the evolution of the Solar System over several hundreds of Myr, we have decided, 
since La2004 \citep{LaskRobu2004a} to use direct integrations, without averaging. 

The dynamical model and numerical integrators are very close to the ones of 
 La2004. We will thus refer to \citep{LaskRobu2004a} for a detailed description of these models, 
 and will only report here the elements that are different in the present 
 model and integration.

\subsection{Dynamical model }
The orbital model  comprises   all 8 planets of the 
Solar system,  and Pluto. The post-Newtonian 
general relativity corrections of order  $1/c^2$ due to the Sun
are included following   \citet{SahaTrem1994a}. 
The Moon is treated as a separate object. In order to 
obtain a realistic evolution of the  Earth-Moon system,
we also  take  into account the most important  coefficient 
in the gravitational potential of the Earth 
and of the Moon, and the tidal dissipation in the Earth-Moon System
\citep{LaskRobu2004a}.  
Contrary to La2004, the precession and obliquity are now integrated without averaging 
over the orbital periods following \citep{BoueLask2006a}. 
In the final runs, we have also added the contribution of 
the main 5 minor planets and some small correction to precession motions are 
made to take into account the remaining asteroids or other unmodelized parameters.

\subsection{Numerical integrator}
As in La2004, the numerical integration was performed with the 
symplectic integrator scheme $SABAC_4$   of \citep{LaskRobu2001a},
with a correction step for the integration of the Moon.
This integrator is particularly adapted to perturbed systems where the 
Hamiltonian governing the equations of motion can be 
written  as the sum of an integrable part  
(the Keplerian equations of the planets orbiting the 
Sun and of the Moon around the Earth), and a small perturbing potential 
representing the interactions among the planets. 

The step size    used in the integration is 
$\tau = 5\times 10^{-3} {\rm yr} =1.82625$ days, while for La2010a, 
$\tau = 10^{-3} {\rm yr} =0.36525$ days.
The initial conditions of the integration were least square adjusted 
to a special version of INPOP that has been extended in time over 1 Myr.
Depending on the solution,  this fit was performed over 
1 Myr or 580 kyr (see Tab. \ref{tab.solutions}).

In La2004, the integration was made in double precision, with machine Epsilon 
$\eps_M \approx 2.22 \times 10^{-16}$. Here, we have  integrated the solutions in extended 
precision on Xeon Intel processors, which allows arithmetics in 80 bits instead of 
64 bits in  double precision. The machine Epsilon becomes then 
$\eps'_M \approx 1.1 \times 10^{-19}$. 

The integration 
time for our complete model, including 5 asteroids and the Moon as a separate object
with $\tau = 5\times 10^{-3} {\rm yr}$
is about  one day per 3 Myr in extended precision, and one day per 6 Myr in double precision
on a Intel Xeon E5462 2.8 Ghz workstation.
When the step size is decreased to $\tau =  10^{-3} {\rm yr}$, for the nominal solution 
La2010a, the requested time is 5 days for 3 Myr, and more than one year for the whole integration. 

\subsection{Numerical  error }

As in La2004, the numerical error is estimated by comparing two integrations 
with the same model and slightly different step size. For the nominal solution, we use 
$\tau = 5\times 10^{-3}$ yr, and for the alternate solution 
 $\tau^* = 4.8828125 \times 10^{-3}$ years. 
This special value is chosen 
in order that our output time span $h=1000$ years  corresponds to an integer number (204800) 
of steps, in order to avoid any interpolation problems in the check of the numerical accuracy.
With  $\tau =  10^{-3} {\rm yr}$, we have then $\tau^* =  0.9765625 \times 10^{-3} {\rm yr}$
(Tab. \ref{tab.solutions}).

\subsubsection{Rotational evolution}

Contrarily to La2004, the precession equations are no longer averaged over 
the orbital motion of the planets or the Moon, but are treated in 
a vectorial manner, following \citep{BoueLask2006a}.
 Following  \citep{Darw1880a,Mign1979a}, we assume that
the torque resulting from tidal friction is proportional to the time lag
$\dt$ needed for the deformation to reach the equilibrium. 
This time lag  is supposed to be constant, and 
the angle between the direction of the tide--raising body and 
the direction of the high tide (which is
carried out of the former by the rotation of the Earth)
is proportional to the speed of rotation. 
Such a model is called ``viscous'', and corresponds to the case for which $1/Q$ is proportional to
the tidal frequency.

Various additional small dissipative effects as core--mantle friction
\citep{Poin1910a,Roch1976a,LumbAldr1991a,CorrLask2003b}, 
atmospheric tides \citep{ChapLind1970a,Voll1978a,CorrLask2003a}, 
mantle convection \citep{FortMitr1997a}, climate friction 
\citep{Rubi1990a,Rubi1995a,Bill1994a,ItoMasu1995a,LevrLask2003a}, 
have been discussed in La2004, but their effects are 
considered to be too small and too uncertain to be added in the model, as it 
was the case for La2004.

\section{The numerical ephemeris INPOP}
\llabel{sec.inpop}

The initial conditions of La2004 were obtained by adjustment to the JPL numerical ephemeris
DE406 \citep{Stan1998b} over the full range of DE406, that is 
from $-5000$ yr to $+ 1000$ yr from the present date. DE406 is itself adjusted to 
planetary observations. 

With this procedure, we are limited by the range of the available ephemeris, and 
in general, the latest ephemeris is not always computed over a long time interval. 
For example, the  most recent ephemeris from JPL, DE421 \citep{Folk2008a}, has only been provided over 
the time interval $[1900, 2050]$ yr. Moreover, it is difficult to estimate the 
true uncertainty of the provided ephemeris. Most often, this  uncertainty is only revealed 
with the publication of the new ephemeris that can be the compared with the previous one. 
In order to overcome these limitations, we have undertaken in our group 
the construction of a full size planetary and lunar ephemeris. After five years of 
work, the solutions are now mature and  two successive versions have already been 
published : INPOP06 \citep{FienManc2008a} and INPOP08 \citep{FienLask2009a}.
The detailed information about the dynamical models and fit to available observations can 
be found in the related publications. 

We have removed in the construction of these ephemerides 
all elements that would limit the length of validity of the solutions. 
In particular, we have not used some precession formulas for the evolution of the spin axis of the Earth. 
Instead, we have integrated together with the full ephemeris, a  precession model 
for the Earth that is obtained after averaging over the rotation period of the Earth, but not over 
the orbital period of the Earth or of the Moon \citep{FienManc2008a}.

The full ephemeris could then be prolongated over 1 Myr using extended precision 80 bits 
arithmetics with the Adams integrator of INPOP. This integration took about 4 month of CPU 
on an itanium 9040 1.6 Ghz workstation. This process was first made for INPOP06, and then for INPOP08, 
when the final version of this latest ephemeris \citep{FienLask2009a} was finally 
made available.
These highly accurate ephemerides are then used for the calibration and evaluation of the long
models La2010.

\tabNa

We refer to \citep{FienManc2008a,FienLask2009a} for a precise description of INPOP06 and INPOP08. 
With respect to INPOP06, INPOP08 benefitted 
from several additional sets of observations. The Mars Express and Venus Express ranging data provided 
very precise measures of Earth--Mars and Earth--Venus distances with a precision of a few meters 
\citep{FienLask2009a}. For Mars, this was a continuation of a long sequence of very precise 
measures that had been acquired with the Martian spacecrafts since the first Viking landers on Mars, 
but for Venus, the new ranging data processed by ESOC were the first 
highly accurate estimates of the Earth-Venus distance which uncertainty was thus reduced from 
a few hundred meters to a few meters \citep{FienLask2009a}.
Another improvement in INPOP08 consists in the use of some Cassini 
normal points \citep{Folk2008a} that also help to constrain  the position of  Saturn. 
In addition, in INPOP08, the Lunar orbit was fitted  to Lunar laser ranging data in  a 
consistent way, while in INPOP06, the fit of the Lunar ephemeris was only made 
with respect to Lunar distances given by DE405  \citep{Stan1998b}.

It is always difficult to estimate the true uncertainty of an ephemeris. In \citep{FienLask2009a}, 
this estimate is obtained by comparison with INPOP06 and  DE421 over 10 and 100 years. 
The differences between INPOP08 and DE421 are in general smaller, but comparable  with 
the differences INPOP08-INPOP06, with the notable exception of the positions of Saturn, 
where the differences INPOP08-DE421 are one order of magnitude smaller than the differences 
INPOP08-INPOP06. This is certainly the consequence of the use in both DE421 and INPOP08 of the 
new Cassini data that constrain very much the position of Saturn. 

After 100 yr, the differences INPOP08-DE421 in barycentric positions range from a few kilometers for the 
inner planets to a few thousands km for the outer planets, and only 40 km for Saturn \citep[][Tab. 6]{FienLask2009a}. 
This is several order of magnitude more than the error coming from the numerical integration 
\citep[][Tab. 1]{FienManc2008a}  that reach only a few micrometers after the same range, and less than a few meters 
after 10000 yr. It can thus be assumed that after one million year, the numerical error in the integration of 
INPOP will still be smaller than the propagation of the uncertainty of the model and 
parameters, obtained by the fit to planetary positions. 

We have thus prolongated the two INPOP ephemeris (INPOP06 and INPOP08) over 1 Myr in order to use 
these solutions as a starting point for the long term ephemeris. 
The accuracy of these solutions after 1 Myr is then evaluated by the comparison of INPOP06 to INPOP08, 
with the assumption that the real uncertainty of INPOP08 will be smaller than the difference INPOP08-INPOP06
(Table \ref{tab.inpop}). In Table \ref{tab.inpop}, we also provide the differences of two 
integrations of INPOP08 made in double precision and in extended precision, in order to 
evaluate the numerical precision of the integration. 
From the comparisons made in \citep{FienManc2008a}, we can assume that the error on the integration of INPOP08 
made in extended precision is in fact several orders of magnitude smaller than the  differences reported in 
this table.

\section{Successive versions of La2010}

\tabNi

The process leading to a long term solution is long, as we had to wait first for the INPOP solution to be ready 
over 1 Myr, and then only we could make  the fit of the long term model. 
After  that, the integration of the long term model over 250 Myr in extended precision  still required 
about 3 months of CPU time, and more than one year 
when the step size is reduced to $10^{-3}$ yr. 
This is why we have performed  several versions of these long term ephemeris, 
that could be used for comparisons, and also to study the stability of the solution with respect 
to improvement of the INPOP ephemeris. As the INPOP08 ephemeris was only finished very recently, 
some of the solutions fitted to INPOP08 have been fitted over only 580 kyr  instead of 1 Myr for INPOP06, 
but this did not make a large difference, and  the solutions are still at the end compared 
to INPOP08 over 1Myr as  the integration of INPOP08 has now reached 1Myr. 
The various models that have  been selected are summarized in Table \ref{tab.solutions}.

\figNe

The solution La2010d has been fitted to INPOP06 over  1 Myr, while the more recent solutions 
 La2010a,  La2010b,  La2010c, and their associated solutions 
 La2010a*,  La2010b*,  La2010c*, have been fitted over INPOP08, over 580 kyr for 
 the solutions with index a,b, and over 1 Myr for La2010c and La2010c*.
All these models, except La2010c and La2010c*
comprise the five major asteroids, Ceres, Vesta, Pallas, Iris and Bamberga. 
 
In all cases, the parameters are taken from the corresponding INPOP ephemeris, as well as 
the starting value of the initial 
conditions. In order to take into account the differences of models, we then perform a fit 
of the semi major axis, and add a small precessing term that can be thought as representative of 
the  average contribution of the minor planets that have not been taken into account 
in our simplified models.
In these solutions, the Moon is integrated as a separate object, taking into account the 
tidal dissipation in the Earth--Moon system. The step size is then $5\times 10^{-3}$ yr
for La2010b,c,d and $ 10^{-3}$ yr for the nominal solution La2010a.

In order to check the numerical accuracy of the solution, we have also integrated these 
solutions with an alternate step size of  $\tau^*= 4.8828125 \times 10^{-3}$ yr for b and c, and 
$\tau^*=0.9765625 \times 10^{-3}$ yr forLa2010a*. 
These values are different, but close to the nominal step size. They are taken 
in such way that $1024 \times \tau^* = \tau$, where $\tau$ is the nominal step size 
of the corresponding solution.
In Fig.\ref{FigNe}, the difference in the eccentricity of the Earth for two solutions 
La2010x and La2010x*  are plotted over time for 100 Myr for $x=a,b,c$. 
Because  of the exponential divergence of the solutions resulting from chaotic behavior, 
the difference is nearly zero for a very long time, of more than 50 Myr, and then 
grows rapidly to maximal value, as the two solutions will become out of phase. 

This is an external way to evaluate the precision of the numerical integration, 
but it is in fact a pessimistic view. 
Indeed, we have fitted the solutions  La2010a,b,c to INPOP08, but  the initial conditions
of  La2010x* is the same as for La2010x. 
The difference of step size will then induce a difference of reference Hamiltonian in the 
symplectic integration of the system, which should explain most of the difference that is observed here. 
This is why for a reduced step size, as in La2010a,  the difference between La2010a 
and La2010a* is smaller.

Nevertheless, although pessimistic, this shows that the numerical error can be neglected 
over 55 Myr for La2010b,c and 60 Myr for La2010a. This is why we have selected 
La2010a as our nominal solution. 

\figNa

\tabNb
\tabNc

\section{Comparison with INPOP08}
\llabel{sec.compinpop}

The solution La2004 was fitted to DE406 over its full range, that is over the interval 
$[-5000:+1000]$ yr from now. In 2004, there was no possibility to compare it to an accurate ephemeris 
over a longer time. With the construction of the new INPOP ephemerides, this becomes 
now possible, as we have extended INPOP06 and INPOP08 over 1 Myr. 
As the set of observation used in INPOP08 is significantly larger than the one of INPOP06, 
we will use INPOP08 as the reference ephemeris, representing the best knowledge of the orbital 
motion of the Solar System that we can achieve at present. 

We have thus compared La2004  to INPOP08 as well as INPOP06 and  the new computed solutions
of the Earth eccentricity for
La2010a,b,c,d (Fig. \ref{FigNa}). From Fig. \ref{FigNa} (top), it is clear that 
the new solution La2010a is a significant improvement with respect to La2004. Indeed, the 
difference  La2010a-INPOP08 (Fig. \ref{FigNa} (a3) and (b3))
is nearly 15 times smaller than the difference La2004-INPOP08 (Fig. \ref{FigNa} (a1)).

In Fig. \ref{FigNa}, 
we can  see that  La2010d-INPOP08 (a2) is almost superposed with  INPOP06-INPOP08 (a4). 
This is because La2010d has been adjusted to INPOP06. It is also clear from this plot that
the differences between a long time solution and its reference ephemeris 
are now much smaller than those of two consecutive versions of the high resolution planetary ephemeris 
(as INPOP06 and INPOP08). 

The main result of these comparisons, are also displayed in Tables \ref{tab.Nb} and \ref{tab.Nc}, 
for  the various solutions that we have selected. 
The differences between  the long term  ''simplified'' model 
that we use here  and the most precise  planetary ephemerides  are  now much 
smaller than the difference between two consecutive planetary ephemeris (INPOP06-INPOP08) 
that can be considered as representative of the true uncertainty of  the ephemeris. 
The main limitation of the precision of the long term  planetary solution 
then resides in the precision of the planetary ephemeris, that 
is in the  planetary observations.

\section{Comparison with La2004}
\llabel{sec.compLa04}
\figNb
\figNc
After comparing the solutions over 1 Myr with the most precise planetary ephemeris, 
we will now  compare the various solutions La2010a,b,c,d to the former  solution La2004 
over the whole expected range of validity of these solutions, 
that is over a few tens of million of 
years. 

In (Fig.\ref{FigNb}), we have represented the variation of the eccentricity of the EMB 
from $-30$ Myr to $-50$ Myr for the La2004 solution and the four  La2010a,b,c,d new solutions. 
The  interval $[-30:0]$ Myr is  not represented as all five solutions practically 
coincide over this time interval. Indeed, some discrepancies between La2004 and the new 
La2010 solutions  only  appear on the time interval $[-40:-30]$, although 
most of the time the solutions  are still very similar, and the small differences 
that can be seen on Fig.\ref{FigNb} will most probably  not lead to any 
significant change in the paleoclimate records. 
We can even consider that the solution  La2004 is still in good agreement with the new solutions 
until $-45$ Myr. This is in good agreement with the expected precision that 
was forecasted in \citep{LaskRobu2004a}.

Beyond $-45$ Myr, noticeable differences become to appear, and the solution La2004 
becomes significantly different than the La2010 solutions. We have thus 
made an additional plot of the $[-50:-40]$ Myr interval on Fig.\ref{FigNb}, with only 
the solutions La2010a,b,c,d. These latest solutions well agree on this time interval, despite the variations 
of models or initial conditions among these new solutions.
We  can thus consider that the present new solutions are at least valid over 50 Myr.

Beyond $-50$ Myr, the situation is more confused as the solutions La2010a,b,c,d 
present significant variations  (Fig. \ref{FigNc}). Nevertheless, from 
Fig. \ref{FigNc}, it can be seen that if moderate precision is only required, 
the solution could eventually used up to $-60$ Myr. In particular, the solutions 
are still well in phase around $-55$ Myr. 
This date is of particular interest, as it corresponds to specific climatic events that have 
been well documented in the paleoclimate records : the Paleocene-Eocene Thermal Maximum
(PETM) that is dated  around 55.53--56.33 Ma\footnote{Myr design the duration 
of 1 million of years, while Ma stands for mega-annum, and represent date in the past 
from the present.}
\citep{WestRohl2007a,WestRohl2008a}, and 
the Elmo Thermal Maximum (ETM) dated at 53.7--54.5 Ma 
\citep{LourSlui2005a,WestRohl2007a,WestRohl2008a}.

On the other hand, from Fig. \ref{FigNc}, it is quite clear that the solutions
La2010a,b,c,d cannot be used beyond $60$ Ma, as the solutions have large differences 
in the interval $[-65:-60]$ Myr. It should thus be stressed that 
in this time interval, the  direct use of  the eccentricity solution of the Earth 
for  geological calibration should be used with  utmost care. 

Practically, 
if one wishes to use these solutions beyond 50 Myr,  for a geological calibration 
for example, one should use only the features of the solutions that 
remain the same in the four La2010 solutions. This will assess 
the stability of such a calibration with respect to 
 the uncertainty of the La2010 solutions. 

\section{Long term cycles}

\tabd
\figNf
\figNda
\figNd
\tabc

The complete eccentricity  solution of the Earth allow a direct adjustment of 
paleoclimate data to the  oscillations of about  95 kyr and 124 kyr of 
the eccentricity \citep[see][]{LaskRobu2004a}, but  for
ancient records, this signal may not be  clearly visible in the sediments. 
On the other way, the 405 kyr  oscillation   with argument $g_2-g_5$, where 
$g_i$ (Table \ref{tab.freqsec}) are  the  secular frequencies of the Solar System 
\citep[see][]{LaskRobu2004a} is very often  present in the sedimentary records. 
This term is the largest term in a quasi periodic approximation of the eccentricity of 
the Earth \citep[see][Table 6]{LaskRobu2004a}, and is less 
influenced by the chaotic  diffusion present in the Solar system than the 
shorter period terms around 100 kyr \citep{Lask1990a,LaskRobu2004a}.

In recent works, the modulation of the 405 kyr component, due to the beat 
$g_3-g_4$ of period $\approx$ 2.4 Myr has also been identified 
in the sedimentary records, and  is  thought to be a key factor 
for the onset of special climate events \citep{LourSlui2005a,PaliNorr2006a,DamAziz2006a}. 
More generally, there has been a large effort to search 
for long term cycles in the sedimentary records and 
to use them to hook the  sedimentary data to the eccentricity computations 
\citep{OlseKent1996a,LourSlui2005a,WestRohl2007a,WestRohl2008a,JovaSpro2010a}.
In figure \ref{FigNf}a is plotted the spectrum of the nominal solution La2010a, 
limited to the range $[0,5]$ "/yr (periods larger than 260 kyr). As there is a gap at
about $2.2$ "/yr in this spectrum,  we have  filtered  the  eccentricity data 
for all various solutions in  the range $[0,2.2]$ "/yr (Figures \ref{FigNda}, \ref{FigNd}).

Here again, it is clear that all solutions La2004, and La2010a,b,c,d are  practically 
identical over $[-30:0]$ Myr, and very similar up to - 45 Myr, where La2004 starts to differ 
notably from the new solutions La2010, which still behave in a similar  manner 
up to about -50 Myr where the situation becomes more confused.

\subsection{The modulation of the $g_2-g_5$ 405 kyr cycle}
In order to examine more closely  the long term cycles in the Earth eccentricity, 
we have identified  the origin of the main spectral terms in the  eccentricity spectrum
of figure \ref{FigNf}a. In order to do so, a synthetic  eccentricity curve is 
build along the same time range using only the five terms  of 
$e\exp(i \varpi)$, as provided by the frequency decomposition of the solution 
La2004 over the time interval $[-15,+5]$ Myr,  taken from \citep{LaskRobu2004a}.
The plot of the spectrum of the eccentricity function that is obtained with this purely 
quasiperiodic signal with frequencies limited to the linear terms 
$g_1,g_2,g_3,g_4,g_5$ (Table \ref{tab.zecc}) is  plotted in 
Figure \ref{FigNf}b.

As this synthetic model is quasiperiodic  with only five main frequencies, 
the identification of the  main spectral terms of the eccentricity 
are then obtained unambiguously  with a spectral analysis over 65 Myr 
and are detailed in figure \ref{FigNf}b.
These terms are easily related to the corresponding  peaks of the 
full eccentricity spectrum of Figure \ref{FigNf}a.

This exercise, that is in some sense complementary from a  full quasiperiodic decomposition 
of the eccentricity as in \citep[][Table 6]{LaskRobu2004a}  allows to better understand 
the behavior and origin of the  main long term cycles observed by the practitioners 
when comparing to geological data 
\citep{OlseKent1996a,LourSlui2005a,WestRohl2007a,WestRohl2008a,JovaSpro2010a,HilgKuip2010a}.

The leading periodic term is  the well known 405 kyr term $g_2-g_5$, but this term is surrounded 
by two terms $(g_2-g_5)-(g_4-g_3)$ and  $(g_2-g_5)+(g_4-g_3)$  that will induce 
with $g_2-g_5$  a modulation of the 405 kyr eccentricity term  with  a frequency of $g_4-g_3$, 
corresponding to a 2.4 Myr period. 
An obvious consequence is that when analyzing geological data to search for the 405 kyr term, 
one needs to use a spectral window that includes these two side terms, that is a window 
similar to the $[2.2,4.3]$ "/yr window used in figure \ref{FigNg}.
Additionally, the two terms $g_1-g_5$  of period $\approx 1$ Myr, and 
$g_2-g_1$ of period $\approx 688$ kyr are also of strong amplitude in the eccentricity 
spectrum. 

\subsection{The $g_4-g_3$ 2.4 Myr cycle}

\figNg
\figNgg

Moreover,  $g_4-g_3$ appears also directly as a main periodic term of the eccentricity 
(Fig.\ref{FigNf}a,b). The  same 2.4 Myr cycle can thus be directly retrieved from 
the eccentricity curve.  Indeed, in  figure \ref{FigNg}, we have plotted both the 
filtered eccentricity in the interval  $[2.2,4.3]$ "/yr ($e^{a}$) (in red)  and as well the 
filtered eccentricity  with a $[0,0.1]$"/yr window   ($e^b$) (in blue).  It can then be seen that 
the envelope (in green) of $e^a$  is almost identical to the opposite of $e^b$ 
(Fig. \ref{FigNg}).

As a consequence, the two components $e^a$ and $e^b$ of the eccentricity need to be 
added in order to really  evaluate the  component of the  2.4 Myr term (Fig. \ref{FigNgg}).
In the resulting $e^a+ e^b$ curve, the variation of the  maxima are then  attenuated while 
the minima  variations are increased to about 0.02. The variations of the minima are  in phase with the 
$g_4-g_3$ term (plotted in blue in Fig. \ref{FigNgg}).

The large size of these variations makes  it then understandable that 
a signature of these variations could be recorded in the sedimentary  paleoclimate signal.
Indeed, although the global mean annual insolation on earth  varies as $e^2$ and thus 
is not much influenced by eccentricity variations, this is not the case for seasonal variations. 
Indeed, if one considers a black body  with uniform temperature at distance $d$ of 
a star, using Stefan's law for the emission of a black body, 
one finds that its surface temperature $T$ is proportional to $d^{-1/2}$.
In this case, the difference $\delta T$ between perihelion and aphelion temperature will be 
given by 
\be
\Frac{\delta T}{T} \approx \Frac{1}{2}\Frac{2ae} {a}Ê= e \ .
\ee
A change of $0.02$ in the eccentricity corresponds thus in this simplify model  to a change of 
about $0.02\times 300 = 6 $K in the difference between perihelion and aphelion temperatures.

\figNh

Because of the increasing importance of this 2.4 Myr component in  some of the analysis of 
sedimentary records, we have  added here a detailed comparison of the filtered solution 
in the $[0,1.1]$"/yr  interval in figure \ref{FigNh} for the time intervals  $[-55,-40]$ 
and $[-65,-50]$ Myr time intervals. It should be noted that it is not necessary  to compare 
the various  orbital solutions in the $[-40,0]$ Myr 
time interval as they are  practically identical in this range. 

As in the previous discussion, we can see that all curves are very similar until $-45$ Myr, 
while La2004 differs significantly  beyond $-45$ Myr. This is why  this solution is no longer plotted on 
the bottom plot of  figure \ref{FigNh}, which is displayed on the $[-65,-50]$ time interval. 
This range is particularly critical, as it corresponds both to the location of the 
PETM (at about $-55$ Myr) and of the K/P boundary (at about $-66$ Myr) \citep{LourSlui2005a,WestRohl2007a,WestRohl2008a}. 
The various solutions begin to differ significantly  beyond $-53$ Myr, but it can  be 
remarked that the two maxima at about $-57.3$ Myr and $- 59.3$ Myr agree for all four La2010 solutions,
although they largely differ   around $-55$ Myr. 
One could thus use these three peaks  in order to attempt to fit  a geological time scale beyond 
-50 Myr, in the $[-60,-50]$ Myr interval. 

In fact, in the La2010a solution, the $g_4-g_3$ argument has a period of about $2\pi/2.664 \approx 2.36$ Myr
in the interval $[-45,0]$ Myr, but beyond $-45$ Myr, this period changes due to chaotic diffusion
(Fig. \ref{FigNi}). As this occurs at the border of the validity range of the solution, it is still 
difficult to be sure of the real behavior of the $g_4-g_3$ argument beyond $-45$ Myr, 
and it will be necessary to confront these data to geological records to confirm the 
behavior of  the solar system  eccentricity solution. 

\figNi

\section{Resonant angles}
\llabel{sec.resonang}
\subsection{Secular resonances}

\figp

The previous discussion demonstrates the importance of the behavior of the 
$g_4-g_3$ argument in the macroscopic aspect of the variations of  the Earth's  eccentricity, 
and thus its possible relation with the past climate on Earth. 

Indeed, using the secular equations,  \cite{Lask1990a,Lask1992a}   demonstrated that  
the  chaotic behavior  of the Solar System arise from multiple 
secular resonances in the  inner Solar System, and  
in particular,  from the critical argument  associated to
\be
\theta=(s_4-s_3)-2(g_4-g_3) 
\ee
where $g_3, g_4$ are related to the precession of the perihelion of the Earth and Mars,
$s_3, s_4$ are related to the precession of the node of the same planets.
This argument 
is presently in  a librational state, but  can evolve in a rotational state, 
and even  move to libration in a new resonance, namely
\be 
(s_4-s_3)-(g_4-g_3) = 0 \ .
\label{eq.sec3}
\ee

The argument $\theta$ as well as the 
other important resonant argument  ($\sigma= (g_1-g_5) -(s_1-s_2)$) that was identified 
 by \cite{Lask1990a,Lask1992a} as  the origin 
  of the chaotic behavior of the inner planets are 
  plotted in figure \ref{Figp} for all solutions La2004, La2010a,b,c,d.
    In all cases, transition from libration to circulation appear around $-50$ Myr, leaving 
  some uncertainty  to the behavior of the solution beyond this date.

\subsection{Searching for some geological evidence of chaos}

\figNj

The transition from libration to circulation 
of the resonant argument related to $\theta=(s_4-s_3)-2(g_4-g_3)$
is directly linked with the chaotic diffusion of the orbital trajectories. 
Searching  in the geological record for  the evidence of such a transition  would thus 
be an observational confirmation of the past evolution of the Solar System. 

As it appears  in the previous sections, it becomes  more and more difficult 
to  obtain  by numerical computations only the date of the first transition from 
libration to circulation for this resonant argument (Fig. \ref{Figp}).
The direct observation of the individual arguments related to $g_3,g_4,s_3,s_4$ is certainly out of reach.
On the other hand, the argument $\theta$ corresponds to a 
$2:1$  resonance between the two secular terms  $g_4-g_3$ and $s_4-s_3$,  both terms 
being present in the sedimentary records.
We have  discussed about the importance of the $g_4-g_3$ beat in the eccentricity  solution.
In a similar way, $s_4-s_3$  appears as a beat of  about 1.2 million of years in the solution of obliquity,
as the result of the beat between the $p+s_4$ and $p+s_3$ components of the obliquity, where 
$p$ is the precession frequency of the axis \citep[see][Fig. 7]{LaskRobu2004a}. 

With the occurrence of these beats, the detection of 
the resonant state  in the geological data becomes possible. Indeed, the modulation of 1.2 Myr 
of the obliquity appears clearly in the spectral analysis of the paleoclimate record
 from Ocean Drilling Program  Site 926, 
\citep{ZachShac2001a}.  Moreover, using the ODP legs 154 and 199, \citet{PaliLask2004a} 
could find some evidence that the critical argument of 
$\theta$ did not show a transition to circulation at 25 Myr, as in La93, but 
remained in  libration over 30 Myr, as in the  La2004 solution.

Searching for  a transition of  the $(s_4-s_3)-2(g_4-g_3)$ resonance to the 
$(s_4-s_3)-(g_4-g_3)$ resonance, as displayed in figure \ref{Figp}) is difficult, 
as it requires  to obtain both a good signal in eccentricity (or precession) and in obliquity. 
It may be more direct to search only for a modulation of the $g_4-g_3$ (or $s_4-s_3$) period, as it appears 
in figure \ref{FigNi}. Indeed, the change in La2010a at $-45 $ Myr of the $g_4-g_3$ slope, from 
a period of about 2.4 Myr   to  a period of about  2 Myr    also reflects  the same chaotic transition. 

In fact, this change can be directly seen in the eccentricity record (Figs \ref{FigNgg}, \ref{FigNh}), 
as it will induce a change of the time interval from two maxima in 
the filtered eccentricity (Figs \ref{FigNh}) from a 2.4 Myr period to a 2 Myr period. If the 
405 kyr $g_2-g_5$ signal is well present in the geological data, this  becomes then a 
macroscopic feature that can be detectable.
Indeed, a local time scale can be established using the 405 kyr signal,  and the modulation 
period of this signal should be the $g_4-g_3$ term. The transition is then obtained 
as a transition from 6 periods per beat  to 5 periods per beats.  
Such geological data could then 
be able to discriminate among the various La2010 solutions (Fig. \ref{FigNj}). It becomes therefore
even more important to search for good  sedimentary sections where the 405 kyr signal 
is well  determined.

\section{Stability of the $g_2-g_5$ 405 kyr cycle}
\llabel{sec.mesozoic}
\figr
As it was stressed above, 
the  $g_2-g_5$  405 kyr argument has a particular importance in long time geological calibration, 
as it is present in  many sedimentary records and its good stability  \citep{Lask1990a} 
can allow to use it as a reference time scale 
This argument  is indeed visible in many sedimentary records 
of the Early Mesozoic \cite[][and references therein]{OlseKent1999a}. 
As in \citep{LaskRobu2004a}, 
we have tested the stability of this argument over 
the full period of our integrations, that is over 250 Myr
by comparison of its evolution on all retained La2010 solutions and La2004  (\ref{Figr}). 
The present values of $g_2-g_5$ do not differ significantly 
from the value  of La2004  \citep{LaskRobu2004a}. 
The frequency $g_2-g_5$ is thus kept to its La2004 value  
\be
\nu_{405} = 3.200 ''/{\rm yr}
\ee
which corresponds exactly to a period  
\be
P_{405} = 405 000 \ {\rm yr}.
\ee

As seen in Fig \ref{Figr}, the maximum deviation obtained by comparing all solutions 
is about $2\pi$ over 250 Myr, which correspond to a full cycle of 405 kyr after 250 Myr, 
as was given already in \citep{LaskRobu2004a}.

\section{Discussion and future work}
\llabel{sec.discuss}

The new orbital solutions of the Earth that are presented here can be used for paleoclimate 
computations over  50 Myr. Beyond that time interval, the precision of the solution 
cannot be guaranteed but we nevertheless provide the solution over 250 Myr on our 
Web site {\it www.imcce.fr/Equipes/ASD/insola/earth/earth.html}
as reference, and for a possible use, with caution, over the 
full Paleogene period ( up to 65 Myr).

In order to allow  practitioners to test the stability of the solution and 
deduced calibration, we have decided to provide  the   four solutions that have been 
discussed here, La2010a,b,c,d,  La2010a being the nominal solution. 

It should be stressed that La2010a is chosen as the nominal solution because of its better numerical 
accuracy, but there is no strong clue that La2010a should behave better than the other ones. 
Before $-50$ Myr, they behave  practically all in the same way. beyond $-50$ Myr, 
the robustness of a fit could be  tested by changing the solutions. 
Alternatively, in presence of convincing geological record, 
one may conclude that one solution is  more probable than another one. 
This will be in some sense  a feedback from geology to  celestial mechanics. 

Contrary to \citep{LaskJout1993b} and \citep{LaskRobu2004a}, 
we have provided here only the eccentricity solution. 
Indeed, although the model for the Earth rotational evolution has been improved,
the main uncertainty linked to the evolution of the tidal dissipative effect 
in the past is still the main unknown parameter for the precession and obliquity evolution, 
and we thus do not believe that  a new  solution would provide more insight 
than La2004, unless  a full analysis of the geophysical effects is made, in 
confrontation  with the geological records, which becomes then out of the scope of the 
present paper. We thus refer  to La2004 for precession and obliquity. 

On the other hand, we plan to release soon a full high precision, non averaged solution 
of the rotation and precession of the Earth over 1 Myr, as computed from the INPOP model.

After the publication of the La2004 solution \citep{LaskRobu2004a}, 
our goal was to search for an improved solution, valid over the full cenozoic era. 
We must say that we have not reached this goal. Although the present solution 
presents a significant improvement with respect to La2004, it is 
only valid over about 50 Myr. 

The main improvement in the present solution was to use  a 1 Myr version of 
the INPOP ephemeris \citep{FienManc2008a,FienLask2009a} in order to fit the initial conditions 
and parameters of the model. As future versions of INPOP will appear
\citep{FienManc2010a},
we  will be able to better evaluate the real accuracy of our model.

At this point, it is still difficult to say wether it will be possible to obtain a 
precise solution over 65 Myr for the eccentricity of the Earth. 
Indeed, in the present solution, we have  used a more complex model, by adding 
 the five main asteroids in the orbital computation, but this added 
 also some  instabilities in the  system, and although we increased the
 numerical accuracy of the algorithm by using extended precision instead of 
 double precision, we did not reach a better numerical accuracy 
 than in La2004. 
 We  intend to improve on this point, and to increase the numerical 
 accuracy of the solution, as we would like to be sure that the numerical 
 precision is not the limiting factor in the final precision of the solution. 
 In order to do so, we will also need to improve at the same time on the 
 speed of the algorithm, as the present version of our nominal solution 
 La2010a took nearly 18 months to complete.

With the present  solution (La2010), we have reached the limit of the observational data, 
and the limit of predictability for a precise solution of the orbital evolution 
of the Earth. We have thus decided  to provided several possible outcomes instead of 
a single one as usual. 
The   solutions La2010a,b,c,d  are all available 
on the Web site {\it www.imcce.fr/Equipes/ASD/insola/earth/earth.html}. 

Practitioners can  thus check which of these solutions best fit their data 
beyond $-50$ Myr. Moreover,   it becomes clear  that the 
 long periodic terms  related to $g_4-g_3$ in the 
 eccentricity, that appears also  as modulation of the 
 the 405 kyr term in the eccentricity, and the equivalent 
 $s_4-s_3$ term in the inclination, 
are some key  macroscopic features of the orbital solution that 
are imprinted  in the  geological record. Their precise recovery 
can thus provide some clue for the past chaotic diffusion of the orbital motion 
of the Earth.

\appendix

\begin{acknowledgements}
This work was supported by ANR-ASTCM. It 
benefited from support from INSU-CNRS, PNP-CNRS, and CS, Paris Observatory.
\end{acknowledgements}

\bibliographystyle{aa}
\bibliography{../BIBLIO/la2010}

\end{document}